\documentclass{jaa}
\usepackage{natbib}
\usepackage[colorlinks=true,citecolor=blue]{hyperref}
\usepackage{amsmath}
\usepackage{graphicx}
\usepackage{soul}       

\usepackage{orcidlink}  

\graphicspath{{./}{figures/}}

\begin{document}\sloppy

\title{Stellar substructures in the Galactic disc and halo: Properties, origins, and evolution}

\author{Deepak\textsuperscript{1,2}\orcidlink{0000-0003-2048-9870}}
\affilOne{\textsuperscript{1}Aryabhatta Research Institute of Observational Sciences, Manora Peak, Nainital 263002, India\\}
\affilTwo{\textsuperscript{2}Indian Institute of Astrophysics, Bangalore 560034, India\\ E-mail: deepak4astro@gmail.com\\}

%
%

\twocolumn[{

\maketitle


\msinfo{1$^{st}$ March 2024}{4$^{th}$ May 2024}

\begin{abstract}

Spatial, kinematic, and orbital properties, along with ages and chemical compositions of the thin disc, thick disc, and various stellar substructures in the halo, are studied based on data from the LAMOST and {\it Gaia} surveys. The star formation in the Galactic thin and thick disc, with peak metallicities of $-0.20$ and $-0.45$ dex, is found to have peaked about 5.5 and 12.5 Gyr ago, respectively. The thin disc is also found to have experienced an initial star formation burst about 12.5 Gyr ago. The pro-grade population Splash and hot-disc (HD), with peak metallicity of about $-0.60$ and $-0.43$, are found to be about 13.03 and 12.21 Gyr old, respectively, with peak eccentricity of 0.70 and 0.35, are understood to be of {\it in situ} origin. The Gaia-Enceladus/Sausage (GE/S), Thamnos and Sequoia, with peak metallicity of about $-1.31$, $-1.36$ and $-1.56$, are found to be about 11.66, 12.89 and 12.18 Gyr old, respectively, and are understood to be remnants of dwarf galaxies merged with the Milky Way. The HD, Splash, and Thamnos are found to have experienced chemical evolution similar to the thick disc while GE/S, Sequoia, and Helmi stream are found to have experienced distinct chemical enrichment of iron and $\alpha$-process elements.

\end{abstract}

}]



\doinum{}
\artcitid{}
\volnum{45}
\pgrange{1--}
\setcounter{page}{1}
\lp{23}  

\section{Introduction} \label{sec:introduction}  

Understanding the formation and evolution of our home galaxy, the Milky Way, has been one of the most challenging problems in astrophysics over the decades. The first and the most significant attempt in this direction was the study by \cite{EggenLyndenSandage1962ApJ...136..748E}, which suggested a monolithic collapse model for the Galaxy. This model was later refined by the study of observations of globular clusters by \cite{SearleZinn1978ApJ...225..357S}, who suggested a bi-modal population in the Galactic halo: the inner and outer halo. The inner halo is understood to be an old and flattened population with slight pro-grade rotation and is supposed to have formed during a dissipative collapse. In contrast, the outer halo consists of a comparatively younger and spherical population,  and supposedly accreted from dwarf galaxies. Since then, significant advances both in computations and observations led to a suggestion that galaxies, on a large scale, are formed through hierarchical clustering, which, in general, conforms to the concordance model of Lambda Cold Dark Matter ($\Lambda$CDM) model \citep{PressSchechter1974ApJ...187..425P,DavisEfstathiou1985ApJ...292..371D}.

Distances are key for deriving accurate kinematic properties of individual stars and are hard to measure. There were very few accurate distance measurements until the advent of the Hipparcos space astrometry mission in the 1990s, which provided astrometry (parallaxes and proper motions) for a couple of hundred thousand stars in the solar neighbourhood with unprecedented accuracy \citep{PerrymanLindegren1997A&A...323L..49P}. Measured distances with derived stellar kinematic motion ($U, V, W$) coupled with high-resolution spectroscopic surveys of controlled samples within the solar neighbourhood revealed the existence of distinct structures within the Galaxy. For example, the disc of the Milky Way was found to have two chemically and kinematically separate components: the thin disc and thick disc \citep[e.g.,][]{RobinHaywoodCreze1996A&A...305..125R, EdvardssonAndersen1993A&A...275..101E, Ojha2001MNRAS.322..426O, ReddyTomkin2003MNRAS.340..304R, ReddyLambert2006MNRAS.367.1329R}. Similarly, studies identified many a smaller clustering or overdensity of stars as streams both in the halo and in the disc. However, due to limited sample, their origins still need to be better understood. Studies also suggested that the halo contains stars of different populations that have different chemical and kinematic properties \citep{NissenSchuster1997A&A...326..751N, NissenSchuster2010A&A...511L..10N, StephensBoesgaard2002AJ....123.1647S, JonsellEdvardsson2005A&A...440..321J, IshigakiChiba2010PASJ...62..143I}. The launch of {\it Gaia} and its data release brought the issue of the origin of sub-structures into sharp focus. Thanks to the {\it Gaia} space mission and its unprecedented accurate astrometry for a much larger sample of stars compared to earlier Hipparcos space astrometry mission, many of the structures in the Galaxy are now well-defined. However, we have yet to understand the precise origin of each of these substructures and how they are related within the Galaxy.

Several other studies, for example \cite{Fulbright2000AJ....120.1841F, StephensBoesgaard2002AJ....123.1647S, GrattonCarretta2003A&A...406..131G, CarolloBeers2007Natur.450.1020C, NissenSchuster2010A&A...511L..10N, NissenSchuster2011A&A...530A..15N, SchusterMoreno2012A&A...538A..21S, HaywoodDiMatteo2018ApJ...863..113H, HayesMajewski2018ApJ...852...49H, HelmiBabusiaux2018Natur.563...85H, BrookKawata2020MNRAS.495.2645B} and references therein, have reported the presence of two different halo populations in the solar neighbourhood. The presence of spatial and kinematical substructures in the Galactic halo are some of the key pieces of evidence for the working of cosmological $\Lambda$CDM model \citep[for example, see recent reviews][]{Helmi2008A&ARv..15..145H, Klement2010A&ARv..18..567K}

The {\it Gaia} collaboration, based on the second data release from the {\it Gaia} satellite, has shown two well-separated main sequences in the colour-magnitude diagram (CMD) for the Galactic halo \citep{GaiaCollaboration_DR2_ObsaHRD2018A&A...616A..10G}. Similarly, based on CMD, ages and metallicity distribution, recent studies \citep[see][]{HaywoodDiMatteo2018ApJ...863..113H, GallartBernardBrook2019NatAs...3..932G} have also found the bi-modal population in the Galactic halo. The recent study \cite{GallartBernardBrook2019NatAs...3..932G} has shown that these two halo populations are approximately of the same age with significantly different average metallicities.  Based on a comparison with results from galactic simulation, the study attributes the {\it in-situ} origin for the metal-rich red sequence and the metal-poor blue sequence to the external origin of accretion of dwarf galaxies.

Based on data from the APOGEE and GALAH survey \citep{MajewskiSchiavon2017AJ....154...94M, DeSilvaFreeman2015MNRAS.449.2604D}, more recent studies like \cite{KoppelmanHelmi2019A&A...631L...9K, MyeongVasiliev2019MNRAS.488.1235M, BelokurovSanders2020MNRAS.494.3880B, HortaSchiavon2021MNRAS.500.1385H, IshigakiHartwig2021MNRAS.506.5410I, BuderLind2022MNRAS.510.2407B} and references therein, have further investigated the existence and origins of various stellar substructure buried in the inner and outer regions of the Galaxy. For example, \cite{HortaSchiavon2021MNRAS.500.1385H}, discovered a metal-poor stellar structure buried in the inner region of the Galaxy (within about 4 kpc from the Galactic center) which is chemically and dynamically distinct from the other metal-rich stellar populations in that region and must have been a remnant of a dwarf galaxy (having a stellar mass of about $5 \times 10^{8} \rm M_\odot$) merged with the Milky Way at earlier times.

To further probe the origin of various substructures of the Galactic halo, a systematic chemo-kinematical study of a larger sample of halo stars with a wide range in metallicity and kinematical distribution along with age estimates is needed. For this purpose, we use the data from {\it Gaia} astrometric survey along the LAMOST survey \citep{ZhaoZhao2012RAA....12..723Z,XiangTingRix2019ApJS..245...34X}, which is comparatively much bigger with better coverage in the metal-poor region compared to the APOGEE and GALAH survey \citep{MajewskiSchiavon2017AJ....154...94M, DeSilvaFreeman2015MNRAS.449.2604D, BuderAsplund2018MNRAS.478.4513B, BuderSharma2021MNRAS.506..150B}. The stellar sample is discussed in Section \ref{sec:DataSample} Selection of the Galactic thin disc, thick disc and halo samples is discussed in Section \ref{sec:HaloandDiscStarsSelection} while Section \ref{Sec:SubstructuresHalo} details the selection of various stellar populations from the Galactic halo sample. The comparison of the colour-magnitude diagrams for all the selected stellar populations is provided in Section \ref{sec:CMD_HaloSubStr} Section \ref{sec:HaloSubStr_Properties} discusses various global properties (like metallicity, kinematics, orbital parameters, and ages) of the thin and thick disc along with various stellar populations in the Galactic halo. Chemical evolutions of the thin disc, thick disc, and halo’s stellar populations are discussed in Section \ref{sec:AlphaFe_DiscHalo} Finally, in Section \ref{Sec:Conclusion}, we conclude the paper with some ideas for further advancing our understanding of the formation and evolution history of the Milky Way.

\section{Data sample}\label{sec:DataSample}

Our aim for this study is to have a large data sample with good astrometry and photometry (including extinction data) and suitable spectroscopic parameters so that the Galactic disc and halo substructures covering a wide range of kinematics can be identified and examined to learn about their various orbital, kinematic and chemical properties to help understand their formation and evolution histories. We started with {\it Gaia} survey second data release (hereafter {\it Gaia} DR2), the largest astrometric catalogue available to date. It provides a five-parameter astrometric solution (i.e., positions on the sky, parallaxes, and proper motions) and photometric data in the {\it Gaia} $G$, BP, and RP-band magnitudes covering wavelength bands of [330-1050], [330-680], and [630-1050] nm, respectively, for about 1.7 billion stars \citep{JordiGaiaPhotometry2010,BrownGaiaDr2Summary2018}. However, the {\it Gaia} DR2 survey provides extinction data only for a small fraction of stars. 
The lack of extinction correction to magnitude and colour of sample stars may lead to a large scatter in the CMD considering halo stars cover a large distance, and many are at lower galactic latitudes (about 20\% and 11\% of the halo stars in the LAMOST survey are in the Galactic plane with $|b| < 30^{\circ}$ and 20$^{\circ}$, respectively).
To mitigate this problem, we combined the {\it Gaia} DR2 catalogue with the catalogue of {\it V}-band extinctions provided by \cite{QueirozAnders2020A&A...638A..76Q} from the Bayesian isochrone-fitting code {\tt StarHorse} for the LAMOST DR5 survey.\footnote{{\tt StarHorse} output catalogue for LAMOST DR5 survey is available at the CDS via anonymous ftp to \url{http://cdsarc.u-strasbg.fr} (ftp://130.79.128.5) or via \url{http://cdsarc.u-strasbg.fr/viz-bin/cat/J/A+A/638/A76}}
{\tt StarHorse} is a {\tt Python} tool that uses Bayesian analysis of spectroscopic, photometric, and astrometric data to infer distances, extinction, ages, and masses of field stars. Source matching between the two catalogues is done using the {\it Gaia} DR2 source identifier provided in the catalogue from {\tt StarHorse}. We then added the spectroscopic data from the LAMOST DR5 survey provided by \cite{XiangTingRix2019ApJS..245...34X}.\footnote{The LAMOST DR5 catalogue is available at \url{http://dr5.lamost.org/doc/vac}.} 
For this, we first restricted the selection of unique stars with the highest SNR by using {\tt UQFLAG  == 1} and then cross-matched the two catalogues based on sky coordinates with a maximum separation of 5 arcsec. (The difference between reported sky coordinated among these two catalogues is found to be less than 1, 2, 3, 4 and 5 arcsec for about 98.95\%, 99.86\%, 99.96\%, 99.98\%, and 99.99\% of stars, respectively.) This results in a sample of about five million unique stars with the highest SNR from the LAMOST DR5. About 99\% and 95\% of stars in the final cross-match are within 1.0 and 0.5 arcsec, respectively. We further restrict the data to good quality spectroscopic measurements with {\tt QFLAG\_CHI2 == good}, {\tt TEFF\_FLAG == 1}, {\tt LOGG\_FLAG == 1} and {\tt FEH\_FLAG == 1}.
The sample is also restricted to stars with standard error in {\it V}-band extinction $\sigma_{A\rm_V} \leq 0.2$.

For good quality astrometric data from the {\it Gaia} survey, we restricted the sample to stars with parallax error ($\sigma_{\Pi}$) of less than 20\%. To retain stars with good quality proper motions along RA and DEC (i.e., PM$_{\rm RA}$ and PM$_{\rm DEC}$), we restricted the sample to stars with $\sigma_{\rm PM}$ $\leq$ 0.2 mas yr$^{-1}$ or 20\%. Data is also filtered to remove stars with bad astrometric parameters by restricting the sample to stars which qualify the following criteria \citep{GaiaCollaboration_DR2_ObsaHRD2018A&A...616A..10G, LindegrenHernandez2018A&A...616A...2L}:
\begin{multline}
\left( \frac{\tt astrometric\_chi2\_al}{({\tt astrometric\_n\_good\_obs\_al} - 5)} \right) \\ < 1.44\  \times\  \max(1, \exp (-0.4 \times ({\tt phot\_g\_mean\_mag}-19.5)))
\end{multline}
where, {\tt astrometric\_chi2\_al} is astrometric goodness-of-fit ($\chi^2$) value in the along-scan (AL) direction, {\tt astrometric\_n\_good\_obs\_al} is number of good AL observations, and and {\tt phot\_g\_mean\_mag} is apparent $G$-band mean magnitude. 
These filters resulted in a final sample of about 3.5 million stars with good-quality astrometric and spectroscopic data and good-quality extinction measurements.

\subsection{Orbital properties of the sample stars}

Properties of stars' orbit about the Galactic center can provide much information about their origin. For estimation of sample stars' orbital parameters, we used {\tt galpy}, a {\tt Python} package for galactic-dynamics calculations \citep{JoBovy_galpy2015ApJS..216...29B,MackerethBovy2018PASP..130k4501M}.\footnote{The {\tt galpy} Code is available at \url{http://github.com/jobovy/galpy}}. We used {\tt galpy}'s version {\tt v1.7}. For each of the sample stars, we integrated the orbit in {\tt McMillan17} potential \citep{McMillan2017MNRAS.465...76M} for 10 Gyr at 1 Myr interval by specifying the observed quantities (like astrometric parameters and RV) as the initial conditions using {\tt radec=True}.
(The choice of the Galactic potential leads to significant differences in the estimated orbital parameters. For example, {\tt MWPotential2014} potential \citep{JoBovy_galpy2015ApJS..216...29B} results in an offset of about 10$^5$ km$^2$ s$^{-2}$ in the estimated orbital energies \citep[see, for example][]{FeuilletFeltzing2020MNRAS.497..109F} along with a large fraction of sample stars with positive energies when compared to the energies based on {\tt McMillan17} potential, which is commonly used in more recent studies.)
For the present calculations, the input astrometric parameters (i.e., stars' spatial positions, proper motions, and parallax) are taken from the {\it Gaia} catalogue, while RV is from the LAMOST survey. The output provides several orbital parameters like pericenter and apocenter radii ($R\rm_{peri}$ and $R\rm_{apo}$, respectively), eccentricity ($e$), the maximum vertical distance from the Galactic plane ($|Z\rm_{max}|$), angular momentum components along $X$, $Y$, $Z$-axis (i.e., $L_X$, $L_Y$, $L_Z$), total orbital energy ($E$), etc., along with the galactic space velocities ($U$, $V$, $W$) for each of the sample star.

\subsection{Ages and masses of the sample stars}

Stellar ages and masses are two other key parameters to understand the origin scenarios of different stellar populations. 
Age and mass are estimated using the {\it qoyllur-quipu} or {\tt q$^2$} code which is a {\tt Python} based package developed by Ivan Ramirez for determining the stellar parameters \citep[see][for more details]{RamirezMelendezBean2014A&A...572A..48R}.\footnote{The {\tt q$^2$} code is available at \url{https://github.com/astroChasqui/q2}} This code provides estimates of stellar ages based on Yonsei-Yale isochrons \citep[see][for more details]{YiDemarque2001Y2Isochrones,KimDemarque2002Y2Isochrones} and combination of input stellar parameters like star's temperature, luminosity, log$g$, metallicity, etc. along with their respective errors. {\tt q$^2$} can be used in two different modes: {\it plx}-mode and {\it logg}-mode. In the case of {\it plx}-mode, the required input stellar parameters are stars' temperature, parallax, apparent {\it V}-band magnitude, and metallicity. In the case of {\it logg}-mode, the needed input stellar parameters are stars' temperature, log$g$, and metallicity. 
For this study, we used the {\tt q$^2$} code in the {\it plx}-mode. The required stars' apparent {\it V}-band magnitudes are calculated from the $Gaia$'s $G$-band magnitudes based on conversion formulas provided by the {\it Gaia} Collaboration \citep[see][]{EvansRiello2018A&A...616A...4E}. Stars' total metallicity, [M/H], and corresponding error $\sigma_{[{\rm M/H}]}$ are estimated by accounting for $\alpha$-enhancement using following expression from \cite{Salaris1993ApJ...414..580S}:
\begin{equation}\label{salaris}
{\rm [M/H]} = {\rm [Fe/H]} + \log{[C\cdot10^{[\alpha/{\rm Fe}]} + (1-C)]}
\end{equation}
\begin{equation}
\sigma_{[{\rm M/H}]} = \sqrt{ \sigma_{{\rm [Fe/H]}}^2 +  \left( \frac{C\cdot10^{[\alpha/{\rm Fe}]} }{C\cdot10^{[\alpha/{\rm Fe}]}+(1-C)} \right)^{2} \cdot \sigma_{[\alpha/{\rm Fe}]}^2 }
\end{equation}
where, $C = 0.638$, [Fe/H] and [$\alpha$/Fe] are taken from the LAMOST survey. $\sigma_{{\rm [Fe/H]}}$ and  $\sigma_{[\alpha/{\rm Fe}]}$ are the SD in [Fe/H] and [$\alpha$/Fe], respectively, are also taken from the LAMOST survey.

Finally, from the output, the stars' mean age and corresponding standard deviation (SD) are used for further study. Our age and mass estimates are more reliable for main sequence (MS) and MS turn-off stars due to comparatively well-spaced isochrones in the main-sequence (MS) and MS turn-off (MST) phases. Hence, our discussion on age and mass is restricted to stars on the MS and MST selected by restricting the sample to stars with {\it Gaia}'s absolute {\it G}-band magnitude larger than three. To further remove stars with dubious ages, we restrict our discussion to stars with standard error in age, $\sigma_{\rm Age} \leq 2$ Gyr. Standard error in the estimated masses ($\sigma_{\rm Mass}$) for about 82\% and 99\% of our sample stars is less than 0.05 and 0.1 M$_\odot$, respectively.

\begin{figure}
\includegraphics[width=0.24\textwidth]{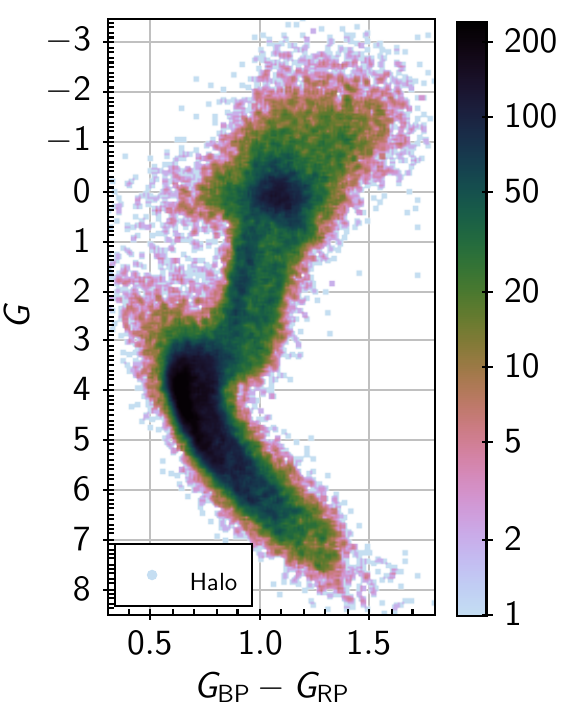}
\includegraphics[width=0.24\textwidth]{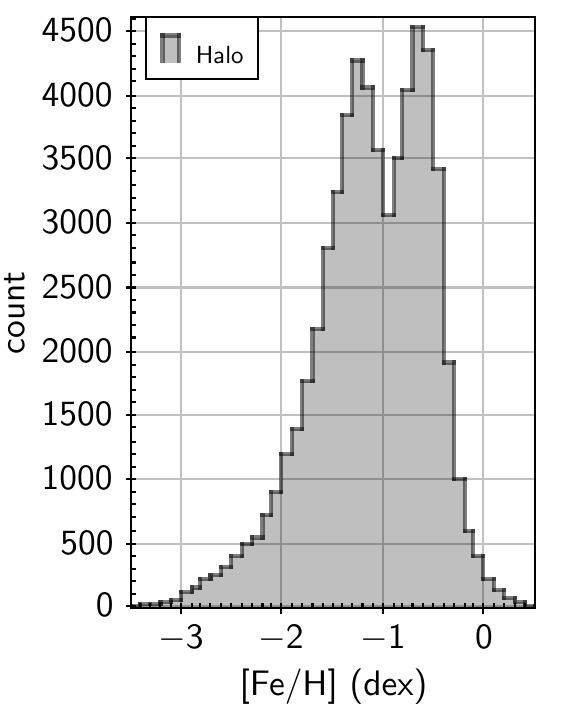}
\caption{Colour-magnitude diagram (CDM) and metallicity function for the selected sample of halo stars. A bi-modal distribution is readily visible in the CMD and metallicity function. In the CMD, for ease in identifying the regions where most of the stars are located, we have shown the number density distributions on a logarithmic scale with the `cubehelix' colour schemes in which the dark black colour represents the highest number density while the light grey colour represents the lowest number density \citep{Green2011BASI...39..289G}.
\label{fig:CMD_Halo}}
\end{figure}

\section{Selection of the Galactic halo and disc stars}\label{sec:HaloandDiscStarsSelection}

\subsection{The outermost stellar component of the Galaxy: Halo} 
From our initial sample of about 3.28 million stars (see Section \ref{sec:DataSample}), we selected the sample of halo stars with transverse velocity (V$_T$) $\geq$ 200 km s$^{-1}$.\footnote{V$_T$ = (4.74/$\Pi$) $\times$ ($\mu_{\alpha}^2 + \mu_{\beta}^2$)$^{1/2}$, where $\Pi$ is parallax in milli-arc-second, and $\mu_{\alpha}$ and $\mu_{\beta}$ are the proper motions in milli-arc-second per year in the right ascension and declination, respectively.} This resulted in a sample of 59 909 stars (henceforth we call it the `halo' sample).
The kinematically selected halo sample is shown in CMD in the left panel of Figure \ref{fig:CMD_Halo}.
The $G$, $G_{\rm BP}$ and $G_{\rm RP}$ are {\it Gaia}'s absolute $G$, BP and RP band magnitude, and are estimated using 
\begin{align}
G = {\tt phot\_g\_mean\_mag} + 5 + 5\times log(\Pi/1000) - A_G \nonumber \ \ \ \ \ \\
G_{\rm BP} = {\tt phot\_bp\_mean\_mag} + 5 + 5\times log(\Pi/1000) - A_{G_{\rm BP}} \nonumber\\
G_{\rm RP} = {\tt phot\_rp\_mean\_mag} + 5 + 5\times log(\Pi/1000) - A_{G_{\rm RP}}
\end{align}
where, {\tt phot\_g\_mean\_mag}, {\tt phot\_bp\_mean\_mag} and {\tt phot\_rp\_mean\_mag} are {\it Gaia}'s apparent $G$, BP and RP band magnitudes. $\Pi$ is {\it Gaia}'s parallax in milliarcseconds. and $A_G$, $A_{G_{\rm BP}}$ and $A_{G_{\rm RP}}$ are {\it Gaia}'s $G$, BP and RP band extinctions derived from the {\it V}-band extinctions ($A_V$) provided by \cite{QueirozAnders2020A&A...638A..76Q}. For conversion of $A_V$ to {\it Gaia}'s extinctions, we used the recipe introduced in \cite{DanielskiBabusiaux2018A&A...614A..19D} and implemented by \cite{GaiaCollaboration_DR2_ObsaHRD2018A&A...616A..10G}.

The CMD for the halo sample, shown in the left panel of Figure \ref{fig:CMD_Halo}, clearly suggests a bi-model population in both MS and red-giants branch (RGB). Of the two populations, the one on the right side (i.e., with a higher colour index) is known as the `red' sequence while the other one (on the left side with a lower colour index) is known as the `blue' sequence.
Origin of these two colour sequences has been associated with a major merging event -- when the Gaia-Enceladus galaxy merged with the progenitor of the Milky Way -- which occurred earlier during the formation of the Galaxy. The blue sequence halo population is understood to be the remnant of the Gaia-Enceladus galaxy. On the other hand, the red sequence halo population is believed to consist of old Milky Way's disc stars, some of the stars of which gained more energetic orbits during the merging event \citep[see][and references therein]{GallartBernardBrook2019NatAs...3..932G}. 
As shown in the right-hand panel of Figure \ref{fig:CMD_Halo} and also suggested in previous studies like \cite{GallartBernardBrook2019NatAs...3..932G}, the bi-model metallicity distribution is the main reason for the presence of two colour sequences in the CMD. (A more detailed discussion on this is provided in Sections \ref{sec:CMD_HaloSubStr} and \ref{sec:HaloSubStr_Properties}) 
These two sequences are also known to have significant metallicity overlap \citep{GallartBernardBrook2019NatAs...3..932G}. This metallicity overlap, and possible errors in the colour index due to hard-to-accurately measure extinctions make it difficult to separate these two major sequences from the CMD. This also renders the detection of other relatively fainter substructures in the halo. Hence, as discussed in the next section we separate various substructures in the halo based on their kinematic properties.

\subsection{Sample of thin and thick disc stars}\label{sec:Sample_ThinThickDisc}

\begin{figure}[ht!]
\includegraphics[width=0.48\textwidth]{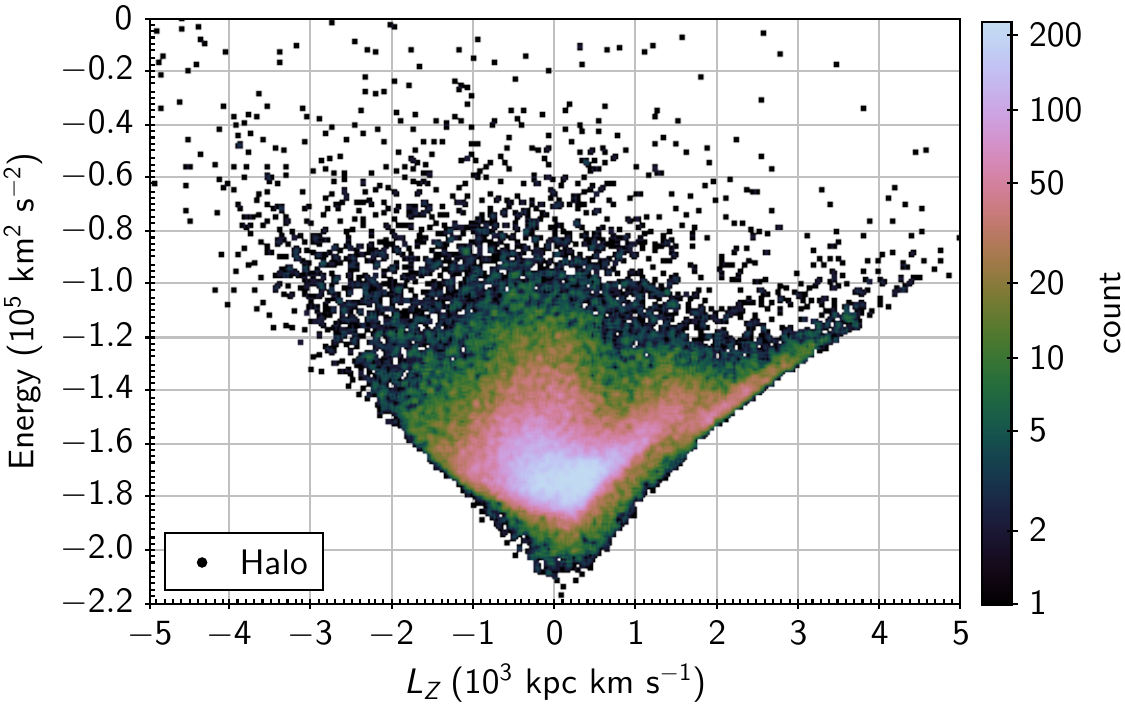}
\includegraphics[width=0.48\textwidth]{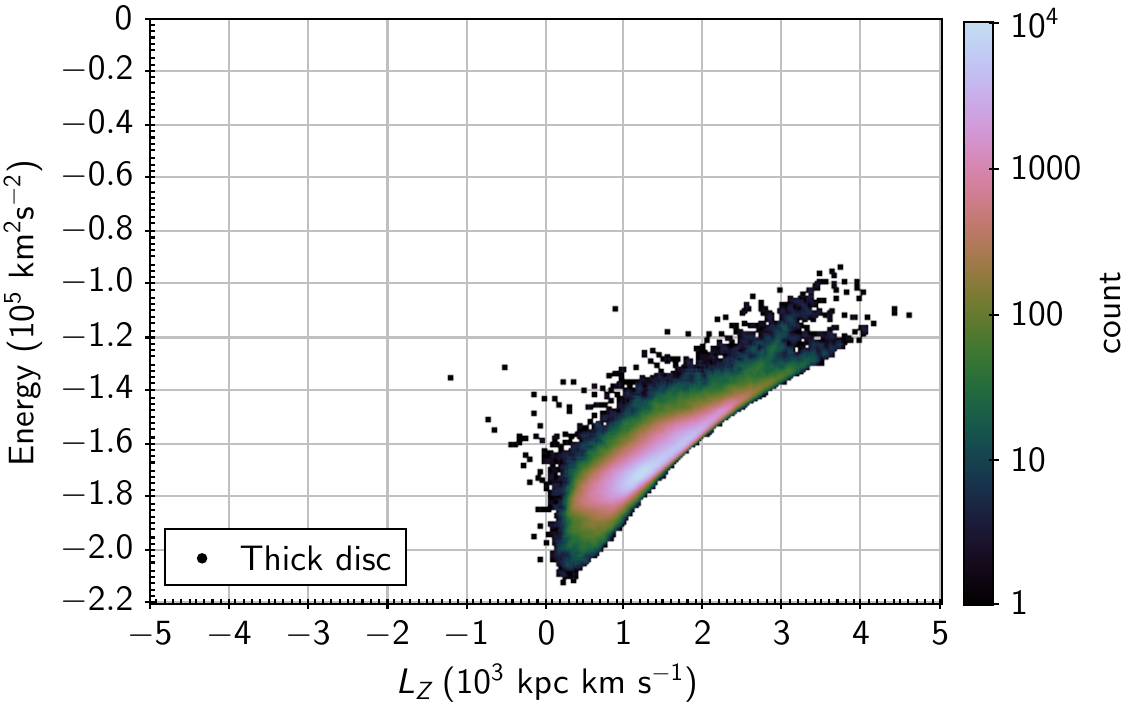}
\includegraphics[width=0.48\textwidth]{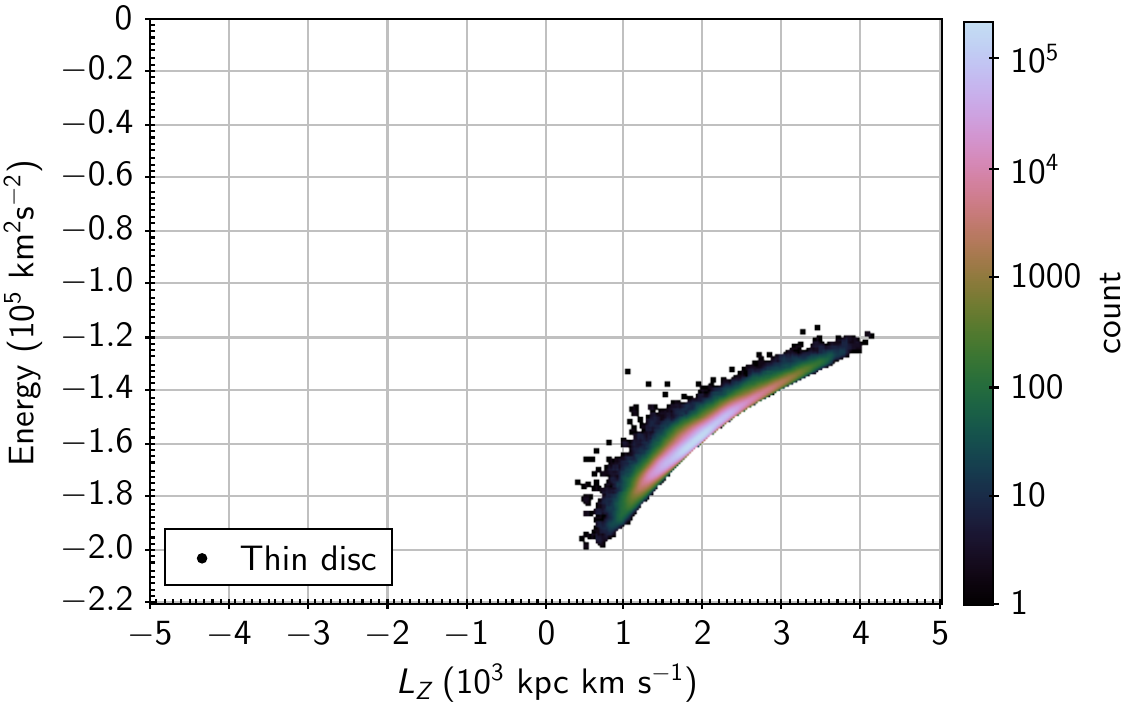}
\caption{Energy-$L_Z$ distribution for the kinetically selected sample of halo, thin disc and thick disc stars. In each of the panels, for ease in identifying the regions where most of the sample stars are located, we have shown the number density distributions on a logarithmic scale with the inverted `cubehelix' colour schemes in which the light grey colour represents the highest number density while the dark black colour represents the lowest number density.
\label{fig:ELz_Halo_ThinThickDisc}}
\end{figure}

To better understand the origin and evolution of the Galactic halo and various components, it is important to compare their properties with the corresponding properties of the thin and thick discs, which are understood to have {\it in situ} origins. For the selection of thin and thin disc stars, we adopt the methodology used in previous studies like \cite{BensbyFeltzing2003A&A...410..527B,ReddyLambert2006MNRAS.367.1329R}. Adopting the kinematic definitions of the thin disc, thick disc and halo from \cite{ReddyLambert2006MNRAS.367.1329R}, we estimate the kinematic membership probabilities to be in the thin disc ($P_{thin}$), thick disc ($P_{thick}$) and halo ($P_{halo}$) for our entire sample of about 3.5 million stars. Finally stars with a minimum membership probability of 80\% to be in the Galactic thin disc are defined as thin-disc stars. Similarly, stars with a minimum membership probability of 80\% to be in the Galactic thick disc are defined as thick-disc stars. This resulted in a sample of about 2.6 million thin-disc stars and about four hundred thousand thick-disc stars, hereafter, we refer to these samples as the `thin disc' and `thick disc' samples, respectively. The distribution of the thin and thick disc sample in the energy ($E$) versus $L_Z$ plane is shown in the bottom two panels of Figure \ref{fig:ELz_Halo_ThinThickDisc}.

\begin{figure*}[ht!]
\includegraphics[width=1\textwidth]{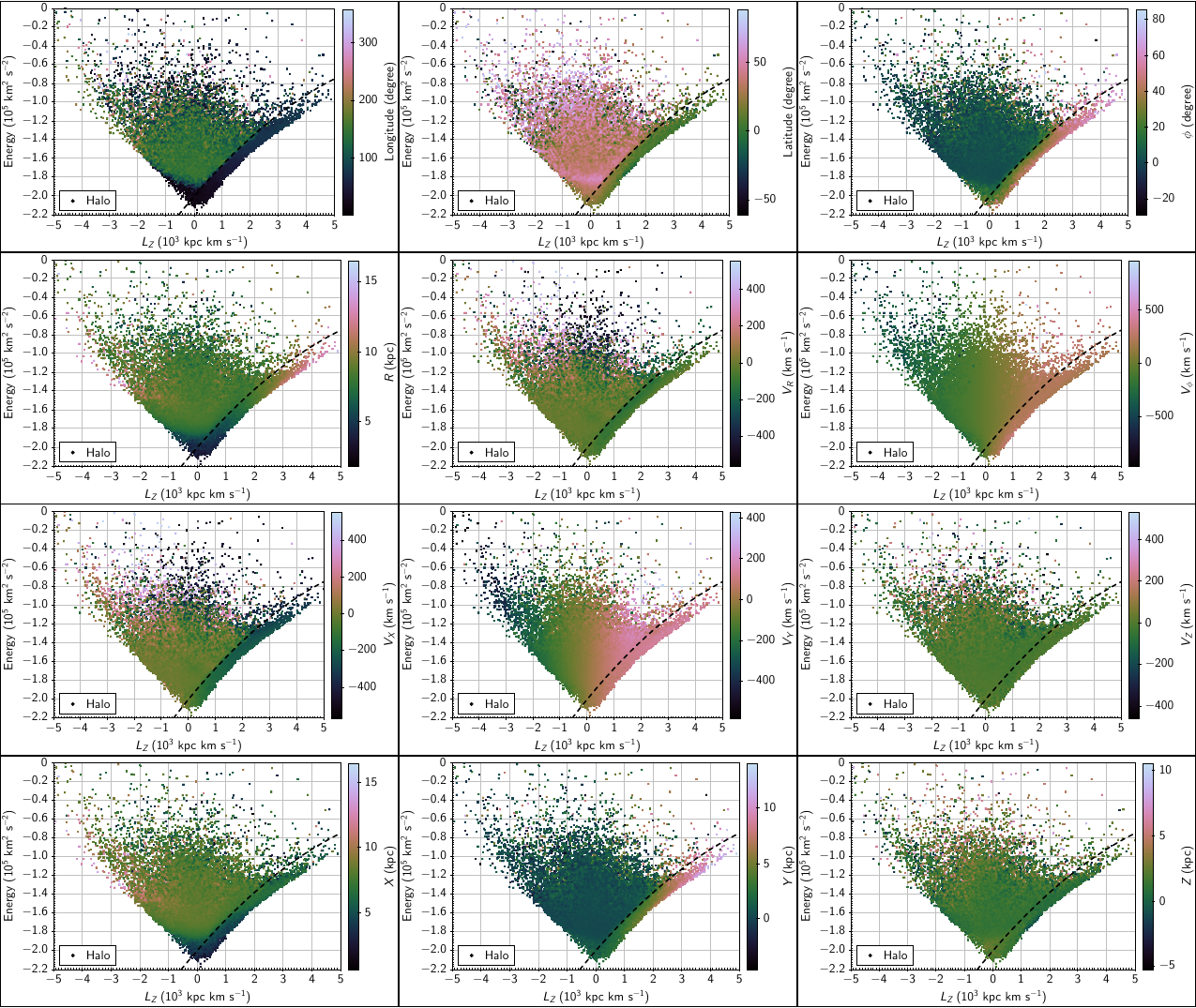}
\caption{Kinetically selected sample of halo stars in $E$-$L_Z$ plane. Stars' sky coordinates, radial distance from the Galactic center $R$, radial velocity $V_R$, $\phi$, and circular velocity $V_\phi$ are shown in colour in the top two panels. Stars' position coordinates ($X, Y, Z$) and space velocities ($V_X, V_Y, V_Z$) are shown in colour in the bottom two panels. The dotted black line represents the quadratic equation $E = -2 \times 10^5 + 35 \times L_Z - 0.002 \times L_Z^2 $ and separates the probable disc stars with large tangential velocity (i.e., stars with $E < -2 \times 10^5 + 35 \times L_Z - 0.002 \times L_Z^2 $) from more certain halo stars (with $E > -2 \times 10^5 + 35 \times L_Z - 0.002 \times L_Z^2 $). For the visualization of density distributions, the used colour scheme is the same as in Figure \ref{fig:ELz_Halo_ThinThickDisc}.
\label{fig:ELz_Halo_WeightLBPhirRxywVxyz}}
\end{figure*}

\section{Stellar substructures in the Galactic halo}\label{Sec:SubstructuresHalo}

Distribution of our kinematically selected sample of halo stars in the orbital energy ($E$) versus angular momentum along $Z$ axis ($L_Z$) plane is shown in the top panel of Figure \ref{fig:ELz_Halo_ThinThickDisc}. A quick comparison of this distribution with previous studies like \cite{HortaSchiavon2021MNRAS.500.1385H} and references therein, suggests the presence of various stellar groups like Gaia-Enceladus/Sausage (GE/S), Thamnos, Sequoia, Halmi stream (HStr), etc., along with disc stars with large tangential velocities. Below we describe the selection of these substructures from the halo sample based on kinematic properties. Unlike \cite{HortaSchiavon2021MNRAS.500.1385H}, we prefer not to make any initial major selection based on the chemical composition of stars considering the LAMOST's low spectral resolution and also to avoid any possible bias in chemical compositions of substructures due to such selection.

\subsection{Disc stars with large tangential velocity}

We start with separating the possible disc stars having high tangential velocities which appear to be buried inside our kinematically selected sample of halo stars. Figure \ref{fig:ELz_Halo_WeightLBPhirRxywVxyz}, shows the kinetically selected sample of halo stars in $E$-$L_Z$ plane with stars' sky coordinates, radial distance from the Galactic center $R$, radial velocity $V_R$, $\phi$, and circular velocity $V_\phi$ in colour in the top two panels while stars' position coordinates ($X, Y, Z$) and space velocities ($V_X, V_Y, V_Z$) in colour in the bottom two panels. Across the range of positive $L_Z$, stars in the lowest energy bands are spatially as well as kinematically different from the rest of the stars in the halo sample. Spatially, these stars are at lower galactic latitude (with $|b|$ $\lesssim$ 30$^\circ$ and peak distribution at $|b|$ $\approx$ 13$^\circ$) and also lower galactic longitude ($l \approx$ 55 $\pm$ 35$^\circ$). Kinematically, as apparent from the $\phi$, $V_\phi$, $R$ and $V_R$ distributions, these stars also have relatively large circular and radial velocities. Same apparent from the distribution of $X, Y, Z$ and $V_X, V_Y, V_Z$ (Figure \ref{fig:ELz_Halo_WeightLBPhirRxywVxyz}). The dotted black line across the panels in Figure \ref{fig:ELz_Halo_WeightLBPhirRxywVxyz} separates these stars from the remaining halo stars. The dotted black line represents the quadratic equation $E = -2 \times 10^5 + 35 \times L_Z - 0.002 \times L_Z^2 $. We classify stars with $E < -2 \times 10^5 + 35 \times L_Z - 0.002 \times L_Z^2 $ as disc stars with large tangential velocity. This resulted in a sample of 4810 stars having disc-like spatial distribution but hotter kinematics (henceforth, we refer to this as the hot-disc sample or in short HD sample). About 60\% of these stars have thick disc membership probability ($P_{thick}$) higher than 50\%, further confirming their kinematic association with the Galactic thick disc.

\begin{figure}
\includegraphics[width=0.48\textwidth]{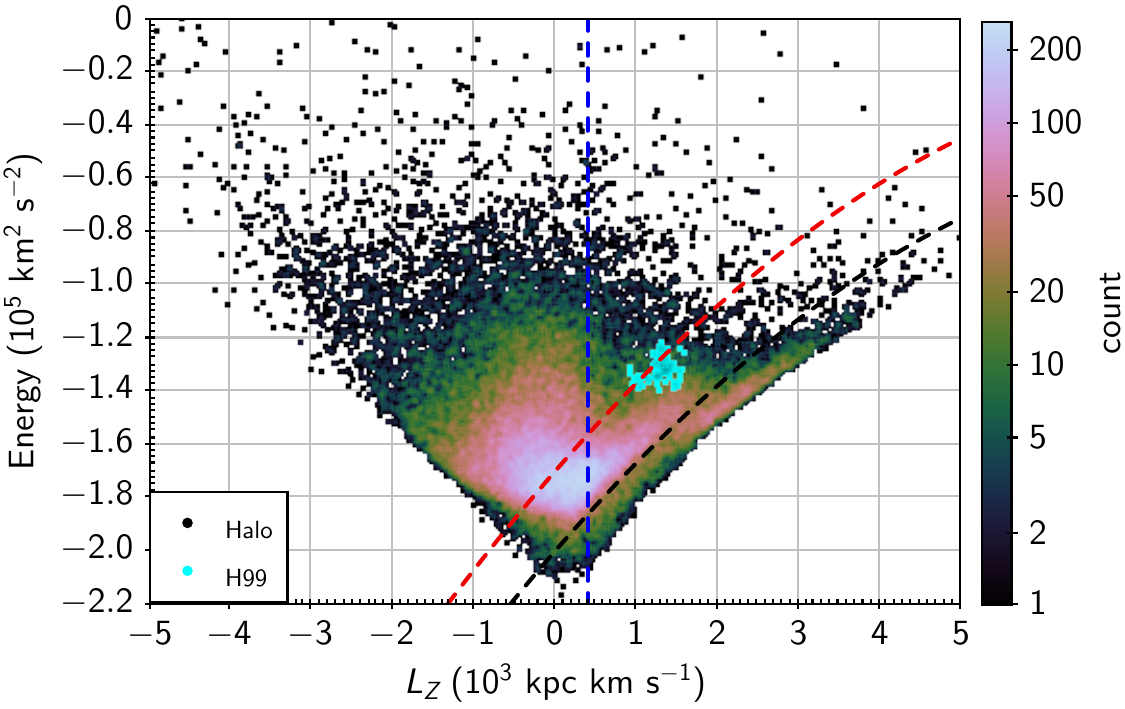}
\caption{Kinetically selected sample of halo stars in $E$-$L_Z$ plane. The dotted black and red lines represent the quadratic equation $E = -2 \times 10^5 + 35 \times L_Z - 0.002 \times L_Z^2 $ and ($E = -1.7 \times 10^5 + 35 \times L_Z - 0.002 \times L_Z^2 $), respectively, and encloses the stars belong to the Splash. Selection of Splash's stars is further restricted to $L_Z$ $>$ 0.4 $\times 10^3$ kpc km s$^{-1}$ to avoid contamination from retro-grade populations. Members of the Helmi stream (H99) are shown in cyon colour. For the visualization of density distributions, the used colour scheme is the same as in Figure \ref{fig:ELz_Halo_ThinThickDisc}.
\label{fig:ELz_SplashSelection}}
\end{figure}

\begin{figure}
\includegraphics[width=0.48\textwidth]{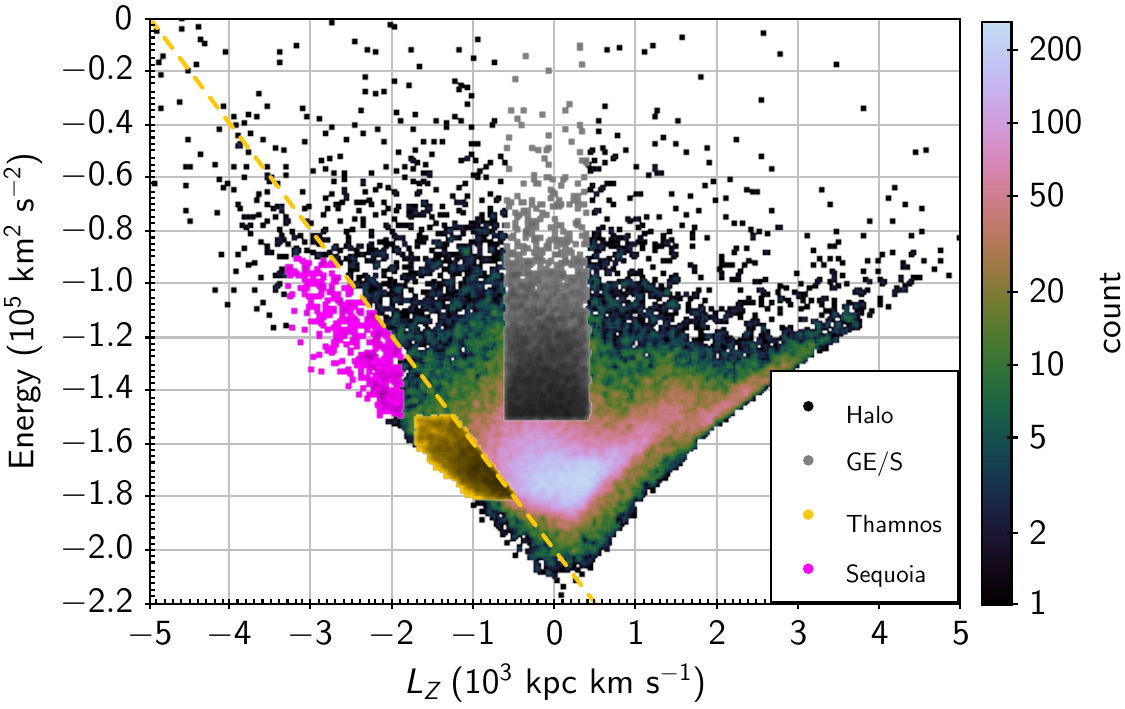}
\caption{Selection of Gaia-Enceladus/Sausage, Thamnos and sequoia. The kinetically selected sample of halo stars is shown in the background.
GE/S stars are selected with $E \gtrsim -1.5 \times 10^5$ km s$^{-2}$ and $-$600 $\leq L_Z \leq$ 400 kpc km s$^{-1}$.
The dotted yellow line represents the quadratic equation
$E < -2 \times 10^5 - 40 \times L_Z$. Stars below the dotted yellow line and with
$-1.8 \times 10^5 < E < -1.5 \times 10^5 $ km$^2$ s$^{-2}$ and $-1700 < L_Z < -500$ kpc km s$^{-1}$ are classified as Thamnos stars, while Sequoia's selection is restricted to stars with $-1.5 \times 10^5 < E < -0.9 \times 10^5 $ km$^2$ s$^{-2}$ and $-3300 < L_Z < -1900$ kpc km s$^{-1}$.
\label{fig:ELz_RetrogradeSubstructures}}
\end{figure}

\subsection{Splash}

As can be seen in Figure \ref{fig:ELz_Halo_ThinThickDisc}, another major pro-grade substructure is present adjacent to the HD component. This substructure is linked to the `Splash' population identified by \cite{BelokurovSanders2020MNRAS.494.3880B}. The selection of the Splash population is shown in Figure \ref{fig:ELz_SplashSelection}. The stars between dotted black line ($E > -2 \times 10^5 + 35 \times L_Z - 0.002 \times L_Z^2 $) and  red line ($E < -1.7 \times 10^5 + 35 \times L_Z - 0.002 \times L_Z^2 $) are potential members of the Splash. However, for $L_Z <$ 0.4 $\times 10^3$ kpc km s$^{-1}$, the Splash population merge with other retro-grade populations in the halo sample. To avoid contamination from the retrograde stars, we restrict our Splash's selection to stars with $L_Z$ $>$ 0.4 $\times 10^3$ kpc km s$^{-1}$. This resulted in a sample of 11577 stars.

\subsection{Helmi stream}

Helmi stream, discovered in \citeyear{HelmiWhite1999Natur.402...53H}, is another pro-grade substructure hidden well inside the halo stars \citep{HelmiWhite1999Natur.402...53H}. For the selection of Helmi stream's stars, we use a 5D kinematic criteria based on inference from previous studies like \cite{HelmiWhite1999Natur.402...53H,LimbergRossi2021ApJ...907...10L}. Selection is restricted to stars with 90 $< V_Y <$ 190 km s$^{-1}$, $-$140000 $< E <$ $-$120000 km$^2$ s$^{-2}$, 900 $< L_Z <$ 1600 kpc km s$^{-1}$ and 0 $< J_R <$ 500 kpc km s$^{-1}$. This resulted in a sample of 93 stars (hereafter, H99 sample). The location of H99 stars in $E$-$L_Z$ plane is shown in Figure \ref{fig:ELz_SplashSelection}.

\subsection{Gaia-Enceladus/Sausage}

Gaia-Enceladus/Sausage is a major accreted component of the halo \citep{MyeongVasiliev2019MNRAS.488.1235M} and readily visible in $E$-$L_Z$ plane with nearly zero angular momentum. At lower energies ($\lesssim -1.6 \times 10^5$ km s$^{-2}$), it overlaps with Splash. However, for $E \gtrsim -1.5 \times 10^5$ km s$^{-2}$, GE/S and Splash stars are well separated. Also, at negative $L_Z$, it further merges with other retro-grade components (like, Thamnos and Sequoia). As shown in Figure \ref{fig:ELz_RetrogradeSubstructures}, we restrict our GE/S sample selection to $E \gtrsim -1.5 \times 10^5$ km s$^{-2}$ and $-$600 $\leq L_Z \leq$ 400 kpc km s$^{-1}$. This resulted in a sample of 6263 stars.

\subsection{Thamnos and Sequoia}

Thamnos and Sequoia are two major highly retrograde components of the Galactic halo \citep{KoppelmanHelmi2019A&A...631L...9K,MyeongVasiliev2019MNRAS.488.1235M,LimbergRossi2021ApJ...907...10L}. Based on inference from the previous studies, we restrict search for Thamnos and Sequoia's to stars with $E < -2 \times 10^5 - 40 \times L_Z$. Further, Thamnos is restricted to stars with $-1.8 \times 10^5 < E < -1.5 \times 10^5 $ km$^2$ s$^{-2}$ and $-1700 < L_Z < -500$ kpc km s$^{-1}$, while Sequoia is restricted to stars with $-1.5 \times 10^5 < E < -0.9 \times 10^5 $ km$^2$ s$^{-2}$ and $-3300 < L_Z < -1900$ kpc km s$^{-1}$. This resulted in 3006 and 306 stars belonging to the Thamnos and Sequoia, respectively. Selection is shown in the Figure \ref{fig:ELz_RetrogradeSubstructures}.

\begin{figure*}[ht!]
\centering
\includegraphics[width=0.8\textwidth]{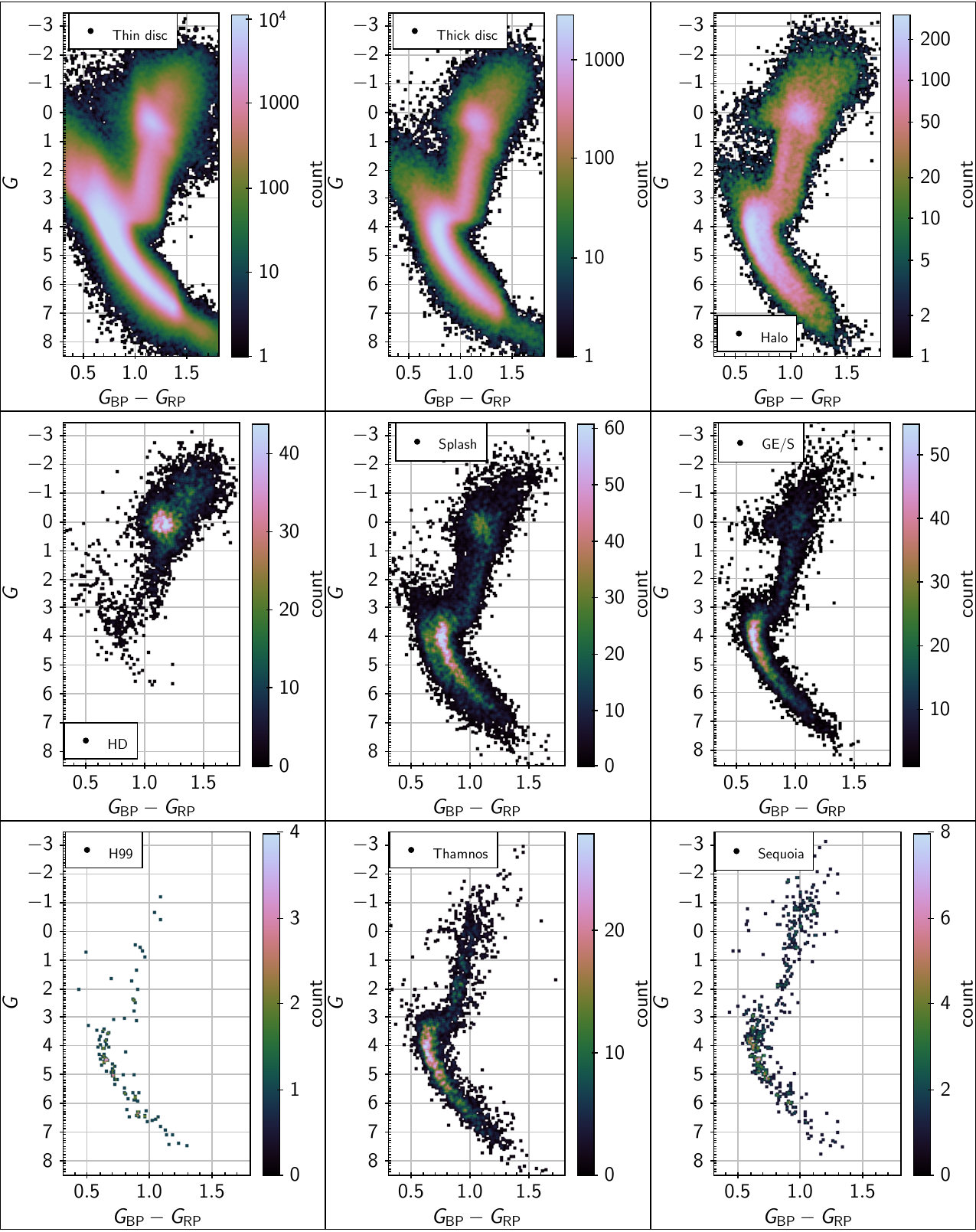}
\caption{Colour-magnitude diagram (CMD) for the thin disc, thick disc and halo sample along with CMDs for all the selected components/substructures of the halo. For the visualization of density distributions, the used colour scheme is the same as in Figure \ref{fig:ELz_Halo_ThinThickDisc}.
\label{fig:CMD_HaloSubStruct}}
\end{figure*}

\section{Colour-magnitude diagrams for the Galactic disc's and halo's stellar populations}\label{sec:CMD_HaloSubStr}

Colour-magnitude diagrams (CMDs) for the thin disc, thick disc and halo, along with the kinematically selected substructures in the halo, are shown in Figure \ref{fig:CMD_HaloSubStruct}. All the disc's and halo's components, except for the HD, have a well-populated MS and RGB. The HD population is dominated mainly by giant stars. For the halo, as shown in the top-right panel of the figure, two colour sequences are readily visible in both MS and RGB. As discussed previously in Section \ref{sec:HaloandDiscStarsSelection}, these two colour sequences correspond to the halo's two major components, the Splash and GE/S. The difference in the metallicities of these two components is the primary reason for the presence of these two colour sequences. As mentioned before, clean segregation of these two sequences purely based on the distribution in the CMD is difficult due to possible errors in the estimated colours and overlap in CMDs due to overlapping metallicities. The overlap in CMDs of the two major components of halo, Splash and GE/S, is evident from corresponding panels in Figure \ref{fig:CMD_HaloSubStruct}, which are based on kinematically selected samples of these two components. Apart from the difficulty in clean segregation of these two major components, the figure also suggests that it would have been impossible to identify other small substructures purely based on the distribution in the CMD, as all these small substructures would have ended up as contaminants among the samples of the two major components, the Splash and the GE/S.

\begin{figure*}[ht!]
\centering
\includegraphics[width=0.88\textwidth]{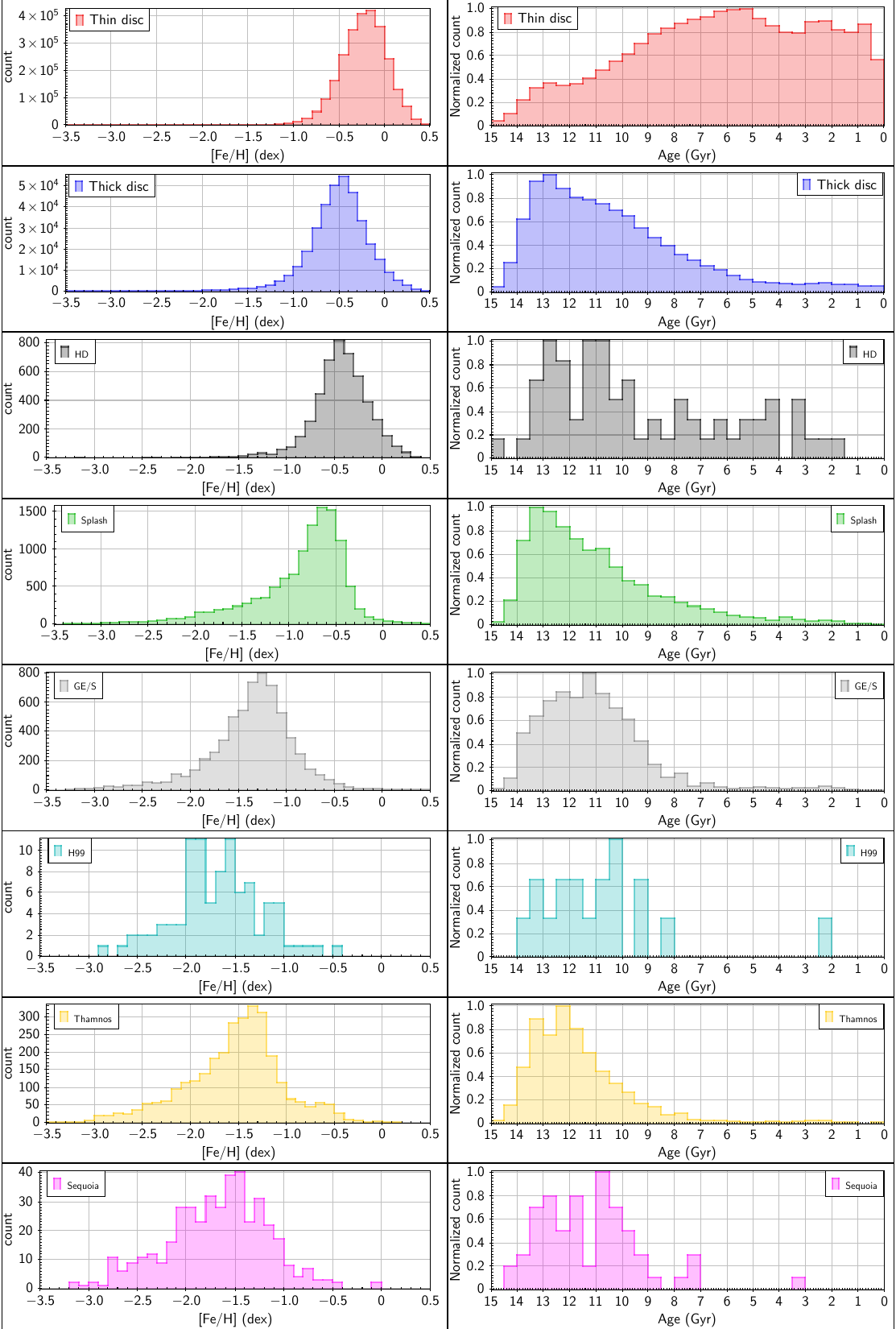}
\caption{Metallicity and age functions for the Galactic disc's and halo's components.
\label{fig:FeHAgeHist_DiscHalo_allSubStru}}
\end{figure*}

\begin{figure*}[ht!]
\centering
\includegraphics[width=0.87\textwidth]{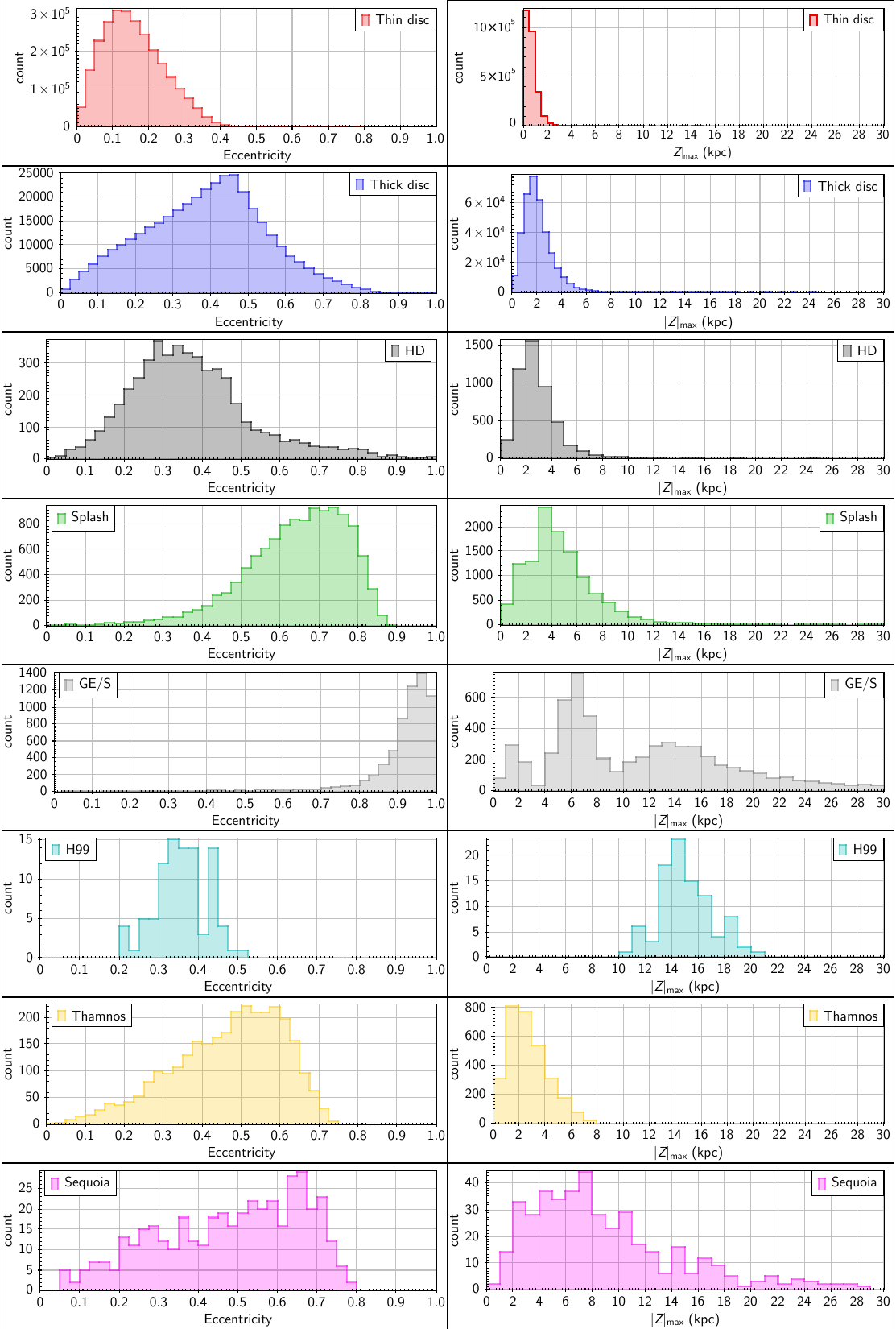}
\caption{Orbital eccentricity and maximum absolute vertical distance from the Galactic plane ($|Z|_{\rm max}$ in kpc) functions for the Galactic disc's and halo's components.
\label{fig:EccentricityAndZmax_Halo_allSubStru}}
\end{figure*}

\section{Metallicity, age and orbital properties of the Galactic disc's and halo's stellar populations}\label{sec:HaloSubStr_Properties}

Of our kinematically selected six halo substructures, HD and Splash, along with the Helmi stream (H99), are pro-grade, i.e. rotating in the same direction as the Galactic disc about the Galactic center (and so also have positive angular momentum $L_Z$). On the other hand, Thamnos and Sequoia are highly retro-grade substructures. The GE/S consists of both pro-grade and retro-grade stars; however, it is also slightly retrograde on average. To find clues about the origins of these substructures, we study the distributions of their kinematic and orbital parameters, age and chemical composition in comparison to the distributions for the thin and thick discs. The metallicity and age functions for all the selected substructures in the Galactic halo along with the thin and thick discs are shown in left- and right-hand panels Figure \ref{fig:FeHAgeHist_DiscHalo_allSubStru}, respectively. Similarly, the left- and right-hand panels of Figure \ref{fig:EccentricityAndZmax_Halo_allSubStru} show the eccentricity and $|Z|_{\rm max}$ functions, respectively.
Statistical parameters derived for all the distribution functions shown in Figure \ref{fig:FeHAgeHist_DiscHalo_allSubStru} and \ref{fig:EccentricityAndZmax_Halo_allSubStru} are summarized in Table \ref{table:FeHAgeEccZmax_1}. 
Further discussions on metallicity, age, and orbital properties for each of the selected stellar populations from the Galactic disc and halo are provided below:

\begin{table*}[ht!]
\caption{Statistical parameters (mean, median, mode, skewness and kurtosis) for distribution functions for all the selected components from the Galactic disc and halo shown in Figure \ref{fig:FeHAgeHist_DiscHalo_allSubStru} and \ref{fig:EccentricityAndZmax_Halo_allSubStru}. Here, Mode values are calculated using the well-known empirical relationship $\rm Mode = 3 \times Median - 2\times Mean$. The table also lists the population standard deviation (SD) and the median of absolute deviations from the median (MAD = $median(abs(x-median(x)$) for each of the stellar populations and provides a robust measure of statistical dispersion for each of the stellar populations. The second column, with the header ``Count" lists the number of sample stars used to estimate the statistical parameters for the corresponding substructure.
\label{table:FeHAgeEccZmax_1}}
\centering
\footnotesize
\begin{tabular}{lrrrrrrrr}
    \hline
 & & \multicolumn{7}{c}{[Fe/H]}\\
\cline{3-9}
  &  & Mean & SD & Median & MAD & Mode & Skew & Kurtosis \\
  & Count & (dex) & (dex) & (dex) & (dex) & (dex) &  & \\
\hline
Thin disc & 2635322 & $-$0.23 & 0.27 & $-$0.22 & 0.17 & $-$0.20 & $-$1.10 & 7.09 \\
Thick disc & 366621 & $-$0.51 & 0.36 & $-$0.49 & 0.19 & $-$0.45 & $-$1.41 & 6.88 \\
HD   & 4810 & { $-$0.43} & 0.30 & $-$0.42 & 0.17 & $-$0.40 & { $-$1.10} & 4.85 \\
HD: low-$\alpha$   & 1643 & { $-$0.22} & 0.24 & $-$0.21 & 0.14 & $-$0.19 & $-$1.18 & 5.06 \\
HD: high-$\alpha$   & 3167 & { $-$0.55} & 0.27 & $-$0.51 & 0.13 & $-$0.43 & $-$1.84 & 8.62 \\
Splash   & 11577 & $-$0.89 & 0.50 & { $-$0.75} & 0.24 & $-$0.47 & { $-$1.31} & 2.15 \\
GE/S   & 6263 & $-$1.37 & 0.45 & { $-$1.31} & 0.23 & $-$1.19 & $-$0.77 & 2.02 \\
Thamnos   & 3006 & $-$1.54 & 0.51 & $-$1.48 & 0.27 & { $-$1.36} & $-$0.41 & 0.72 \\
Sequoia   & 421 & $-$1.68 & 0.53 & $-$1.64 & 0.34 & { $-$1.56} & $-$0.27 & 0.43 \\
H99   & 93 & { $-$1.68} & 0.45 & { $-$1.68} & 0.29 & { $-$1.68} & { 0.08} & 0.12 \\
\hline
 & & \multicolumn{7}{c}{Age} \\
\cline{3-9}
  & & Mean & SD & Median & MAD & Mode & Skew & Kurtosis \\
  & Count  & (Gyr) & (Gyr) & (Gyr) & (Gyr) & (Gyr) &  & \\
\hline
Thin disc & 959647 & 6.11 & 3.59 & 5.91 & 2.80 & 5.51 & 0.27 & $-$0.80 \\
Thick disc & 96876 & 10.32& 2.95 & 10.95 & 1.79 & 12.21 & $-$1.16 & 1.16 \\
HD   & 69 & 9.27 & 3.39 & 10.25 & 2.32 & { 12.21} & { $-$0.60} & $-$0.85 \\
HD: low-$\alpha$   & 9 & 9.36 & 3.26 & 8.70 & 3.05 & 7.38 & $-$0.12 & $-$1.42 \\
HD: high-$\alpha$   & 60 & 9.08 & 3.40 & 10.25 & 2.31 & { 12.59} & { $-$0.52} & $-$0.98 \\
Splash   & 2573 & 10.99 & 2.57 & 11.67 & 1.42 & { 13.03} & { $-$1.33} & 1.71 \\
GE/S   & 1464 & 11.06 & 2.10 & { 11.26} & 1.20 & { 11.66} & $-$1.63 & 4.49 \\
Thamnos   & 765 & 11.60 & 2.02 & 12.03 & 1.01 & { 12.89} & { $-$2.05} & 6.36 \\
Sequoia   & 66 & 11.19 & 1.94 & 11.52 & 1.30 & { 12.18} & { $-$1.20} & 2.83 \\
H99   & 18 & 10.75 & 2.54 & { 11.14} & 1.26 & 11.92 & $-$2.07 & 4.93 \\
\hline
 & & \multicolumn{7}{c}{Eccentricity} \\
\cline{3-9}
  & Count & Mean & SD & Median & MAD & Mode & Skew & Kurtosis \\
\hline
Thin disc & 2635322 & 0.16 & 0.08 & 0.15 & 0.06 & 0.13 & 0.52 & $-$0.20 \\
Thick disc & 366621 & 0.39 & 0.16 & 0.40 & 0.11 & 0.42 & 0.01 & $-$0.38 \\
HD   & 4810 & 0.37 & 0.16 & { 0.35} & 0.09 & 0.31 & { 0.87} & 1.11 \\
HD: low-$\alpha$   & 1643 & 0.32 & 0.12 & { 0.30} & 0.07 & 0.26 & { 0.96} & 2.98 \\
HD: high-$\alpha$   & 3167 & 0.40 & 0.17 & { 0.39} & 0.10 & 0.37 & { 0.68} & 0.55 \\
Splash   & 11577 & 0.64 & 0.13 & { 0.66} & 0.09 & { 0.70} & { $-$0.91} & 1.02 \\
GE/S   & 6263 & 0.91 & 0.11 & { 0.94} & 0.03 & { 1.00} & { $-$3.51} & 16.06 \\
Thamnos   & 3006 & 0.47 & 0.14 & { 0.49} & 0.10 & 0.53 & { $-$0.57} & $-$0.24 \\
Sequoia   & 421 & 0.48 & 0.18 & { 0.51} & 0.14 & 0.57 & { $-$0.42} & $-$0.85 \\
H99   & 93 & { 0.36} & 0.06 & { 0.36} & 0.04 & { 0.36} & { $-$0.19} & $-$0.19 \\
\hline
 & & \multicolumn{7}{c}{$|Z|_{\rm max}$} \\
\cline{3-9}
  & & Mean & SD & Median & MAD & Mode & Skew & Kurtosis \\
  & Count & (kpc) & (kpc) & (kpc) & (kpc) & (kpc) &  & \\
\hline
Thin disc & 2635322 & 0.66 & 0.48 & 0.56 & 0.27 & 0.36 & 1.92 & 9.17 \\
Thick disc & 366621 & 2.12 & 1.19 & 1.92 & 0.66 & 1.52 & 1.75 & 8.86 \\
HD   & 4810 & 2.90 & 1.65 & { 2.58} & 0.84 & 1.94 & { 2.54} & 15.37 \\
HD: low-$\alpha$   & 1643 & 2.50 & 1.62 & 2.10 & 0.76 & { 1.30} & { 2.70} & 14.22 \\
HD: high-$\alpha$   & 3167 & 3.11 & 1.63 & 2.82 & 0.81 & { 2.24} & { 2.65} & 17.48 \\
Splash   & 11577 & 4.73 & 2.82 & 4.22 & 1.47 & { 3.20} & { 1.83} & 8.08 \\
GE/S   & 6263 & { 14.15} & 20.69 & 10.79 & 4.92 & 4.07 & { 16.42} & 503.79 \\
Thamnos   & 3006 & 2.73 & 1.49 & 2.53 & 1.00 & { 2.13} & { 0.73} & 0.08 \\
Sequoia   & 421 & 8.93 & 5.78 & 7.43 & 3.18 & { 4.43} & { 1.29} & 1.69 \\
H99   & 93 & 15.06 & 2.01 & { 14.80} & 1.18 & 14.28 & 0.37 & $-$0.14 \\
\hline
\end{tabular}
\end{table*}

\subsection{Galactic thin and thick disc}

As shown in Figure \ref{fig:FeHAgeHist_DiscHalo_allSubStru}, the thin and thick disc populations significantly overlap in metallicity but have different peak metallicities. The metallicity function for the thin disc peaks at [Fe/H] of about $-$0.20 dex while the thick disc has a peak metallicity of about $-$0.45 (see Figure \ref{fig:FeHAgeHist_DiscHalo_allSubStru} and Table \ref{table:FeHAgeEccZmax_1}). The distribution functions for both populations are negatively skewed with skewness of $-$1.10 and $-$1.41 for the thin and thick discs, respectively.

The thin and thick discs have strikingly different age functions even when the metallicity function for both populations is not very different (Figure \ref{fig:FeHAgeHist_DiscHalo_allSubStru}). Even though both populations span similar age ranges, the thin disc primarily consists of the younger generation of stars while the thick disc consists of older stars. The age function for the thick disc peaked at about 12.5 Gyr ago, and since then, there has been an almost continuous decrease in the fraction of younger stars until about 5 Gyr ago, after which the addition of younger stars to the thick disc is low and almost constant. Similar to the thick disc, the thin disc also has the oldest possible stars in the Galaxy. The age function for the thin disc peaked at about 5.5 Gyr ago, however, unlike the thick disc, the addition of younger stars to the thin disc is relatively very high. The age function for the thin disc also suggests a secondary peak at about 12.5 Gyr ago which coincides with the peak star formation in the thick disc (Figure \ref{fig:FeHAgeHist_DiscHalo_allSubStru}).

The orbital eccentricity distributions for the thin and thick disc stars are shown in the left panels of Figure \ref{fig:EccentricityAndZmax_Halo_allSubStru}. The distribution shows that most thin disc stars have approximately circular orbits. Eccentricity distribution for the thin disc stars peaks at about 0.13 and is slightly skewed towards elliptical orbits with a skewness of 0.52. On the other hand, the thick disc stars have slightly more elliptical orbits than the thin disc stars, although eccentricity for almost all the thick disc stars is less than 0.8. The thick disc's eccentricity distribution peaks at an eccentricity of about 0.42, and has a very small skewness of 0.01. As shown in Figure \ref{fig:Eccentricity_FeH_DiskHalo_allSubStru}, at any given metallicity the spread in eccentricity for the thin disc stars is lower than the thick disc stars. Additionally, on average, metal-poor stars of the thick disc have relatively higher eccentricity compared to the metal-rich stars (see top-right panel of Figure \ref{fig:Eccentricity_FeH_DiskHalo_allSubStru}).

Spatially, as can be seen from the right-hand panels of Figure \ref{fig:EccentricityAndZmax_Halo_allSubStru}, which shows distributions of absolute vertical distance ($|Z|_{\rm max}$) for all the selected stellar populations, both thin and thick disc populations are localized near the Galactic plane. The $|Z|_{\rm max}$ distribution for the thin- and thick-disc populations peaks at 0.36 and 1.52 kpc, respectively, which is in agreement with the relatively hotter kinematics of the Galactic thick-disc stars compared to the thin-disc stars. Additionally, the distribution for both populations is positively skewed with skewness of 1.92 and 1.75 for the thin and thick discs, respectively.

\subsection{HD population}

The majority of stars in the HD population have metallicity [Fe/H] from about $-$1.5 to 0.4 dex with a mean [Fe/H] of $-$0.43 dex and population standard deviation (SD) of 0.30 dex (see Figure \ref{fig:FeHAgeHist_DiscHalo_allSubStru} and Table \ref{table:FeHAgeEccZmax_1}). The distribution is not symmetric and has a small negative skew of $-$1.10. Also, the HD population is metal-rich compared to the thick disc population by about 0.5 dex.

The HD population consist of stars with a wide range in age, with ages ranging from about 2 to 14 Gyr (see corresponding in Figure \ref{fig:FeHAgeHist_DiscHalo_allSubStru}). The HD's age distribution slightly peaks at 12.21 Gyr and is highly skewed towards the younger age (Table \ref{table:FeHAgeEccZmax_1}) and is similar to the age distribution for the thick disc population.

\begin{figure*}
\includegraphics[width=1.00\textwidth]{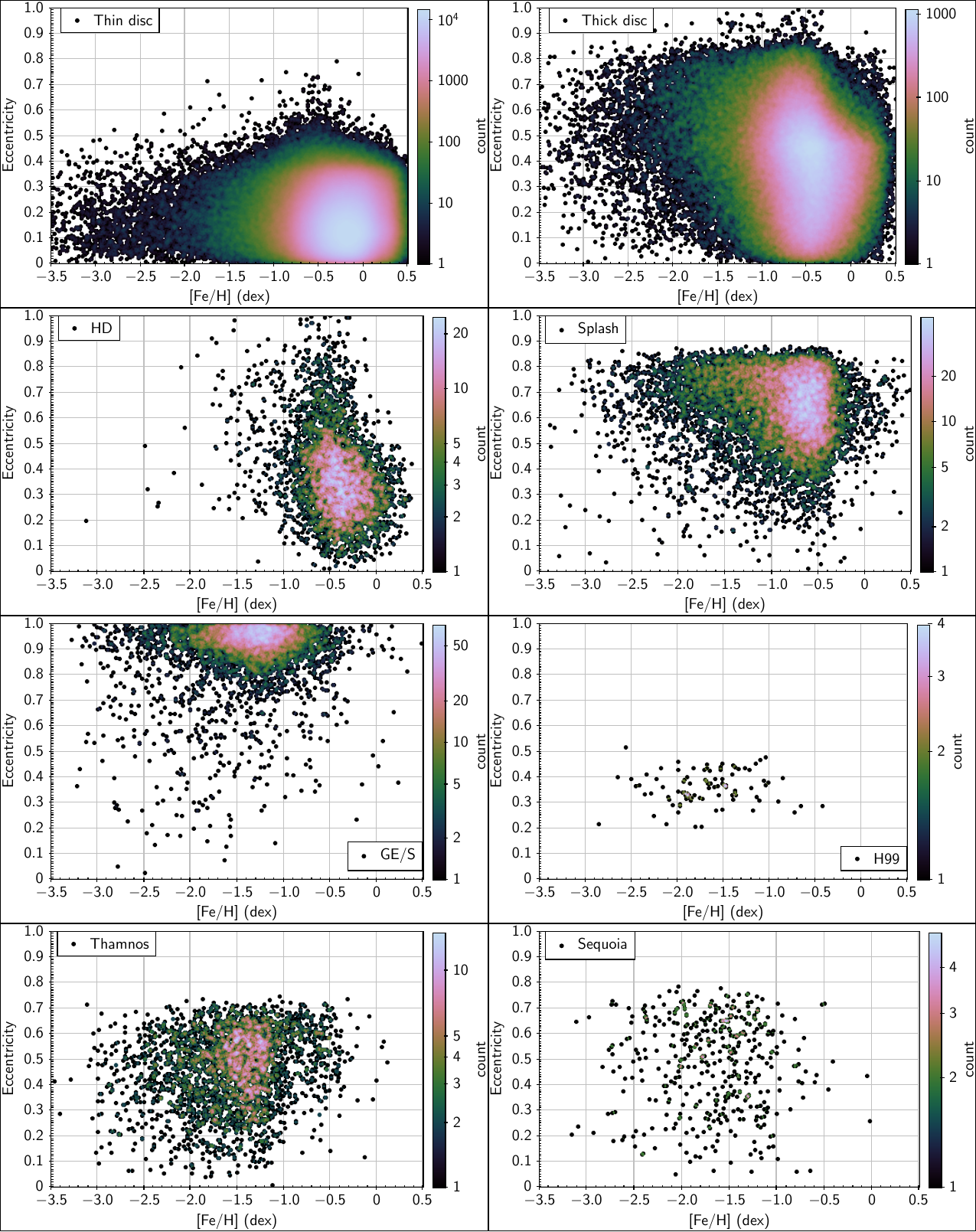}
\caption{Orbital eccentricity as a function of metallicity for all the Galactic disc's and halo's components. For the visualization of density distributions, the used colour scheme is the same as in Figure \ref{fig:ELz_Halo_ThinThickDisc}.
\label{fig:Eccentricity_FeH_DiskHalo_allSubStru}}
\end{figure*}

Kinematically, most of the HD's stars follow a slightly elliptical orbit (Figure \ref{fig:EccentricityAndZmax_Halo_allSubStru}). The sample has a median eccentricity of 0.35 with a median absolute deviation from the median (MAD) of 0.09 (Table \ref{table:FeHAgeEccZmax_1}). However, some of the HD's stars have highly elliptical orbits with eccentricity ranging from about 0.5 to 1.0. Because of this, the distribution is skewed towards the higher eccentricity (Figure \ref{fig:EccentricityAndZmax_Halo_allSubStru}). As shown in Figure \ref{fig:Eccentricity_FeH_DiskHalo_allSubStru}, these high eccentricity stars are slightly metal-poor. The figure also suggests a slightly decreasing trend in eccentricity with an increase in metallicity for HD's stars. The $|Z|_{\rm max}$ distribution for the HD population is asymmetric and has a positive skew of 2.54. Most of the HD's stars are within $|Z|_{\rm max} <$ 6 kpc, and the population has a median $|Z|_{\rm max}$ of 2.58 kpc with a MAD of 0.84 kpc and peaks at about 2 kpc (higher than the corresponding value for the thick disc).

\begin{figure*}[ht!]
\includegraphics[width=1\textwidth]{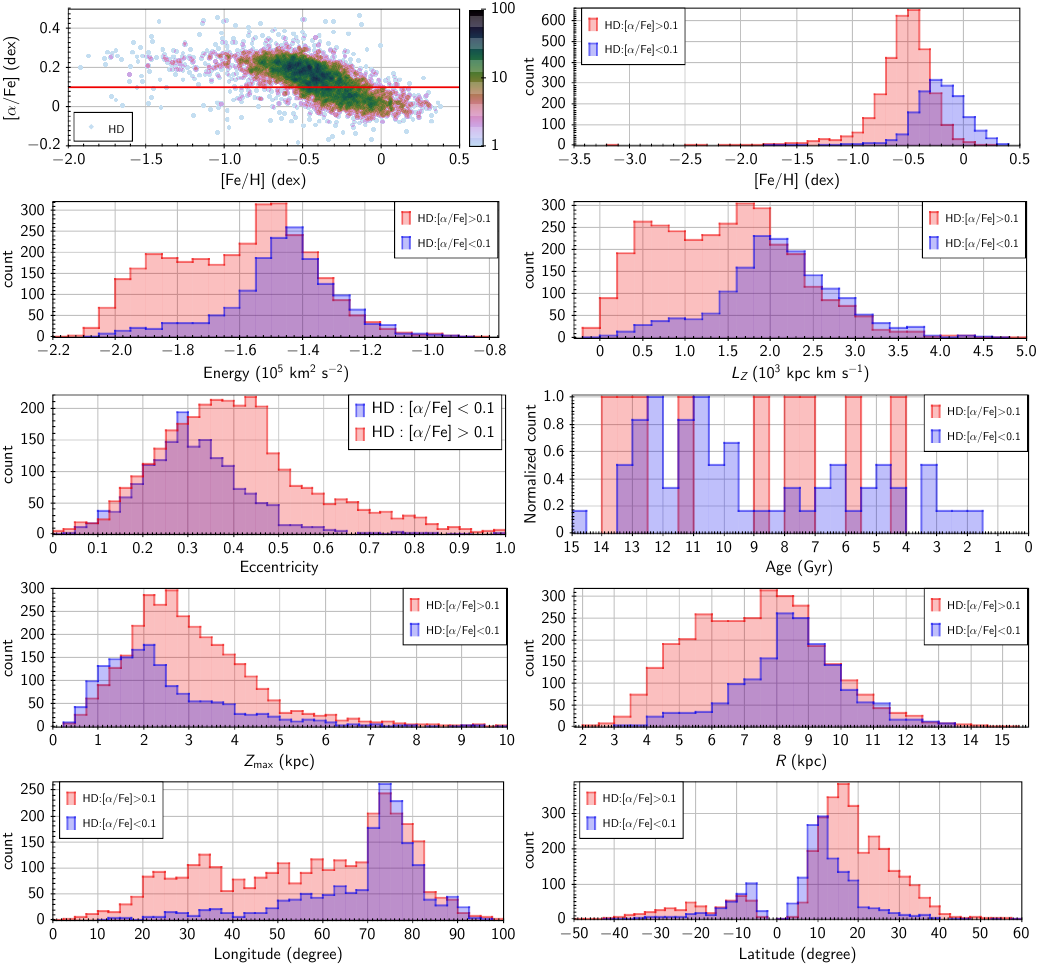}
\caption{[$\alpha$/Fe] versus [Fe/H] distribution for the hot-disc (HD) sample is shown in the top-left panel which also shows the selection of the high- and low-$\alpha$ HD samples. The red horizontal line (with [$\alpha$/Fe] = 0.1 dex) separates the high-$\alpha$ stars (with [$\alpha$/Fe] $>$ 0.1) from the low-$\alpha$ stars having [$\alpha$/Fe] $<$ 0.1. The remaining panels in the figure show histograms for various spatial and kinematic properties along with age distributions for the high- and low-$\alpha$ HD stars.
\label{fig:Hostdisc_alphaFeAndELz}}
\end{figure*}

\subsubsection{Low- and high-$\alpha$ stars in the HD sample:}
The distribution of the HD's stars in the [$\alpha$/Fe] versus [Fe/H] plane is shown in the top-left panel of Figure \ref{fig:Hostdisc_alphaFeAndELz}. This distribution is similar to the distribution of the well-known thick disc \citep[see e.g.,][and reference therein]{ReddyLambert2006MNRAS.367.1329R}. To further understand the properties of the HD, we divide the sample into low- and high-$\alpha$ stars about [$\alpha$/Fe] = 0.1 dex. There are 3167 stars with [$\alpha$/Fe] $>$ 0.1 and 1643 stars with [$\alpha$/Fe] $<$ 0.1 (hereafter, we refer to these as high- and low-$\alpha$ HD samples, respectively). The mean [Fe/H] for HD's low-$\alpha$ population is $-$0.22 with an SD of 0.24, while the mean [Fe/H] for the high-$\alpha$ population is $-$0.55 with an SD of 0.27. This difference in mean metallicity among these two HD's populations is primarily due to how these two populations are segregated based on [$\alpha$/Fe] abundance (see the top-left panel of Figure \ref{fig:Hostdisc_alphaFeAndELz}).

Spatial and kinematic properties along with age functions for the high- and low-$\alpha$ HD samples are shown across the panels in Figure \ref{fig:Hostdisc_alphaFeAndELz}. The low-$\alpha$ HD stars have higher orbital energy ($E$) and angular momentum $L_Z$ than the high-$\alpha$ HD stars, which also span comparatively larger ranges in both $E$ and $L_Z$. The low-$\alpha$ HD stars have comparatively more circular orbits (with peak eccentricity, $e$ $\approx$ 0.3) than the high-$\alpha$ HD stars for which peak $e$ is $\sim$ 0.4 (Figure \ref{fig:Hostdisc_alphaFeAndELz} and Table \ref{table:FeHAgeEccZmax_1}). The age distributions for the low- and high-$\alpha$ populations are similar to the HD population, and the difference, if any, can be attributed to the small sample size. Spatially, low-$\alpha$ HD stars are in the inner region of the Galactic plane (with peak latitude $|b|$ $\sim$ 10$^\circ$) compared to high-$\alpha$ HD stars (with peak $|b|$ $\sim$ 18$^\circ$). Most of the low-$\alpha$ HD stars are at lower $|Z|_{\rm max}$ (with a peak $|Z|_{\rm max}$ of about 1.8 kpc) compared to the high-$\alpha$ HD stars, which have a mean $|Z|_{\rm max}$ of about 2.5 kpc. Additionally, $|Z|_{\rm max}$ distributions for both low-$\alpha$ and high-$\alpha$ HD populations are highly skewed towards the larger value of $|Z|_{\rm max}$ and have a skewness of 2.70 and 2.65, respectively. With respect to the Galactic plane, most of the low-$\alpha$ HD stars are away from the Galactic center and are located at the Galactic longitude $l \sim 75^\circ$. On the other hand, the high-$\alpha$ HD stars are almost uniformly distributed from $l$ $\sim$ 20 to 85$^\circ$, except for a slightly higher accumulation at $l \sim 75^\circ$. The mean radial distance of low-$\alpha$ HD stars from the Galactic center (GC) is about 8.5 kpc, the same as the solar distance. On the other hand, the high-$\alpha$ stars have a mean radial distance of about seven kpc from the GC and span a large radial distance range. In the direction away from the GC, the star count for both low- and high-$\alpha$ samples is similar, but in contrast, towards the GC, the number of high-$\alpha$ is large by many folds. Distribution in $|Z|_{\rm max}$ and radial distance $R$ suggest that the low-$\alpha$ population is more localized in the solar neighbourhood and also restricted to the Galactic disc than the HD's high-$\alpha$ population.

\subsection{Splash}

Splash is one of the major pro-grade components of the Galactic halo. It is also the most metal-rich after the HD population. Its metallicity distribution is highly asymmetric and skewed towards the metal-poor end with a skewness of $-$1.31 (Figure \ref{fig:FeHAgeHist_DiscHalo_allSubStru} and Table \ref{table:FeHAgeEccZmax_1}). The population has a peak metallicity of about $-$0.6 and a median metallicity of $-$0.75 dex with a MAD of 0.24 dex.
Similar to the metallicity, Splash's age distribution is also very asymmetric and highly skewed towards the younger age, with a skewness of $-$1.33. The age distribution peaks at 13.03 Gyr and is highly skewed towards the younger age, with about 50\% of the stars born in about the first 2 Gyr (i.e., from about 14 to 12 Gyr ago) while the majority of the stars from the remaining 50\% are born from about 12 to 5 Gyr ago (Figure \ref{fig:FeHAgeHist_DiscHalo_allSubStru} and Table \ref{table:FeHAgeEccZmax_1}).

Kinematically, the majority of Splash's stars have highly elliptical orbits. Population's orbital eccentricity distribution peaks at 0.70 and has a median value of 0.66 with a MAD of 0.09, and is also highly skewed towards the lower eccentricity with skewness of $-$0.91 (Figure \ref{fig:EccentricityAndZmax_Halo_allSubStru} and Table \ref{table:FeHAgeEccZmax_1}). 
As shown in Figure \ref{fig:Eccentricity_FeH_DiskHalo_allSubStru}, the distribution of eccentricity as a function of [Fe/H] for the Splash is distinct from other populations. The Splash's stars have a maximum eccentricity of about 0.85, and this maximum eccentricity is independent of the star's metallicity. Additionally, the total spread in eccentricity increases with an increase in metallicity. Due to highly eccentric orbits, Splash's stars also reach comparatively larger distances from the Galactic plane. The $|Z|_{\rm max}$ distribution for the Splash peaks at 3.20. However, due to a highly skewed distribution (with skewness of 1.83) towards the higher $|Z|_{\rm max}$, about 16\%, 50\%, and 84\% of Splash's stars have $|Z|_{\rm max }$ lower than 2.2, 4.2, and 7.0, respectively.

\subsection{Gaia-Enceladus/Sausage (GE/S)}

The metallicity distribution function for the GE/S is relatively symmetric (with a skewness of $-$0.77) and has a median [Fe/H] of $-$1.31 with a MAD of 0.23 dex (Figure \ref{fig:FeHAgeHist_DiscHalo_allSubStru} and Table \ref{table:FeHAgeEccZmax_1}).
The age function for the GE/S sample is also relatively more symmetric compared to pro-grade stellar populations discussed before. GE/S's age distribution peaked at about 11.66 Gyr ago and has a median age of 11.26 Gyr with a MAD of 1.20 Gyr.
Also, about 94\% of GE/S stars are older than 8 Gyr, suggesting star formation in GE/S ceased sometime before 8 Gyr ago.

Kinematically, the GE/S stars are on nearly parabolic orbits, with the population having a peak eccentricity of about one and median eccentricity of 0.94 with a MAD of 0.03 (Table \ref{table:FeHAgeEccZmax_1}). Distribution is skewed towards the lower eccentricity with a skewness of $-$3.51.
Additionally, as can be seen in the corresponding panel of Figure \ref{fig:Eccentricity_FeH_DiskHalo_allSubStru}, there is no apparent correlation between orbital eccentricity and [Fe/H] for GE/S stars.
Because of nearly parabolic orbits, the GE/S stars span a wide range in $|Z|_{\rm max}$ ranging from about zero to 30 kpc, and has mean $|Z|_{\rm max}$ of 14.15 kpc with a very large SD of 20.69 kpc, while populations median $|Z|_{\rm max}$ is 10.79 kpc with a MAD of 4.92 kpc. Also, interestingly, the sample suggests three peaks in $|Z|_{\rm max}$ distribution for the GE/S sample (Figure \ref{fig:EccentricityAndZmax_Halo_allSubStru}). These three peaks have peak $|Z|_{\rm max}$ values of about 1.5, 6.5 and 14 kpc. More interestingly, the widths of these peaks also appear to increase with an increase in their peak $|Z|_{\rm max}$. Further studies are required to find if these three peaks are due to the presence of some real kinematic substructures in the GE/S sample.

\subsection{Thamnos and Sequoia}

The Thamnos and Sequoia are the two most metal-poor and highly retro-grade sub-structures in our selected sample of halo stars. Both of these stellar populations span a similar range in metallicity but peak at slightly different metallicities (Figure \ref{fig:FeHAgeHist_DiscHalo_allSubStru}). The metallicity distribution for the Thamnos and Sequoia peaks at $-$1.36 and $-$1.56 dex, respectively (Table \ref{table:FeHAgeEccZmax_1}). These populations also span approximately similar ranges in age; however, their distributions peak at slightly different ages. For Thamnos, the age distribution peaks at 12.89 Gyr, while for Sequoia, it peaks at 12.18 Gyr. Similar to the GE/S, most of the stars in these two populations are older than about 8 Gyr. The presence of a few young stars in the samples can be attributed to either possible contamination in the sample or an error in the age estimation.

Kinematically, stars of both Thamnos and Sequoia have near-circular to highly elliptical orbits. The median eccentricity for the Thamnos and Sequoia are 0.49 and 0.51, respectively, with corresponding MAD values of 0.10 and 0.14. Eccentricity distributions for both populations are skewed towards the lower eccentricity, with Thamnos and Sequoia having nearly the same skewness of $-$0.57 and $-$0.42, respectively (Figure \ref{fig:EccentricityAndZmax_Halo_allSubStru}).
As shown in the bottom panels of Figure \ref{fig:Eccentricity_FeH_DiskHalo_allSubStru}, even with relatively smaller sample sizes, both Thamnos and Sequoia have the largest scatter in eccentricity versus [Fe/H] plane. Also, for both populations, there is no apparent correlation between stars' orbital eccentricity and metallicity.
The two populations have considerable difference in $|Z|_{\rm max}$ distribution (Figure \ref{fig:ELz_RetrogradeSubstructures}). The Sequoia stars span a larger range in $|Z|_{\rm max}$ with a peak $|Z|_{\rm max}$ of 4.43 kpc compared to the Thamnos, which has a peak $|Z|_{\rm max}$ of 2.13 kpc. Also, all the Thamnos stars have $|Z|_{\rm max}<$ 8 kpc, while only about 50\% of the Sequoia's stars have $|Z|_{\rm max}<$ 8 kpc.

\subsection{Helmi stream (H99)}

The metallicity of H99 stars ranges from about $-$2.9 to $-$0.6 dex (Figure \ref{fig:FeHAgeHist_DiscHalo_allSubStru}). Additionally, the H99 population has a nearly symmetric metallicity distribution with a mean of $-$1.68 dex and an SD of 0.45 dex. Our metallicity estimate is higher than mean [Fe/H] = $-2.3 \pm 0.5$, recently reported by \cite{LimbergRossi2021ApJ...907...10L}. One of the main reasons for this disagreement appears to be the relatively small sample in \cite{LimbergRossi2021ApJ...907...10L}, which has only 18 stars compared to our sample of 93 stars. Most of the H99's stars are older than 8 Gyr. The median age of the H99 sample is 11.14 Gyr with a MAD of 1.26 Gyr. 
Kinematically, the H99 stars have slightly elliptical orbits with a mean orbital eccentricity of about $0.36 \pm 0.01$ dex (SD for the H99 population of 93 stars is 0.06 dex), which is in agreement with the mean eccentricity of $0.40 \pm 0.08$ from \cite{LimbergRossi2021ApJ...907...10L}. The median $|Z|_{\rm max}$ for the H99 population is 14.80 kpc with a MAD of 1.18 kpc.

\begin{figure*}[ht!]
\centering
\includegraphics[width=0.98\textwidth]{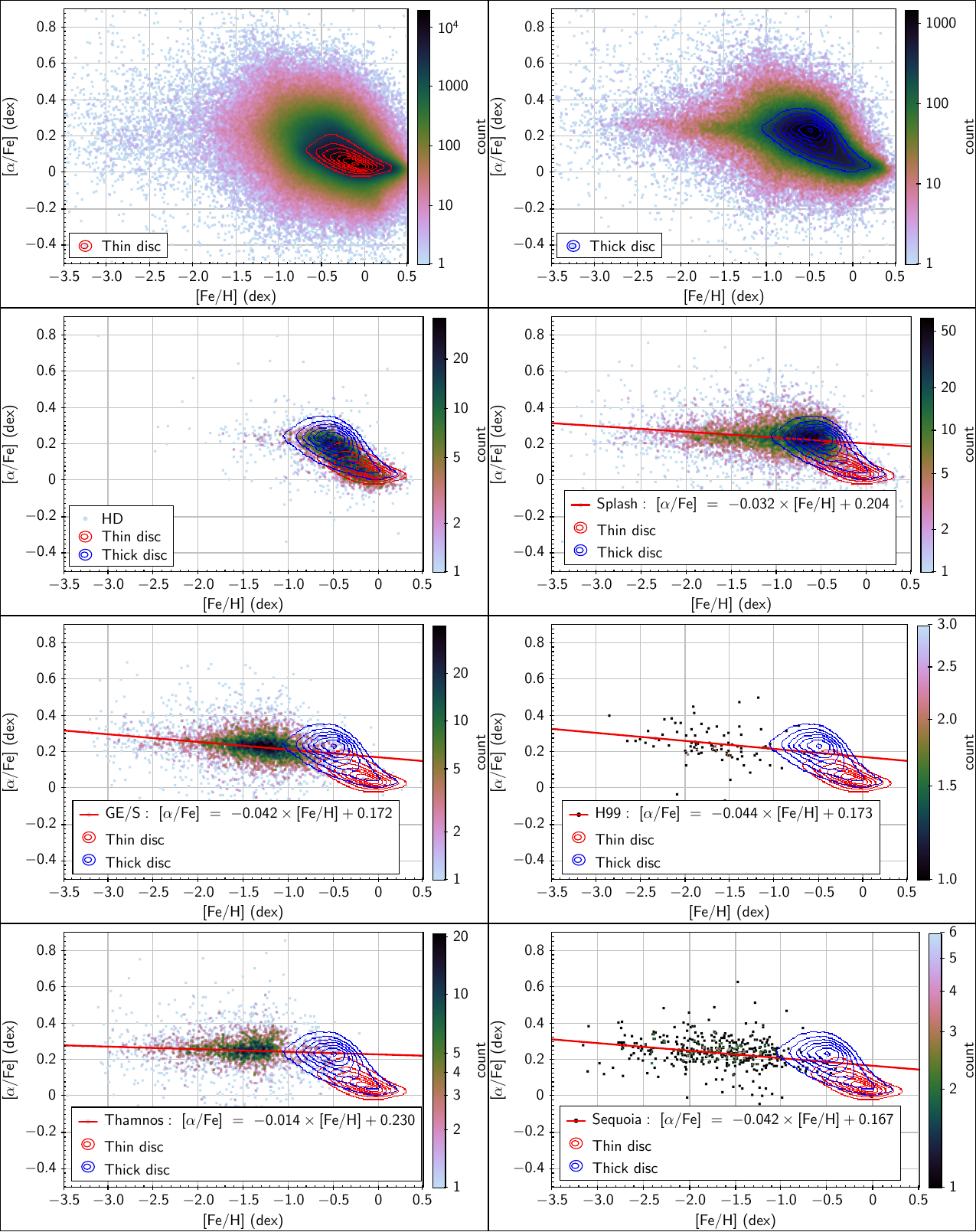}
\caption{[$\alpha$/Fe] versus [Fe/H] distributions for the Galactic disc and halo components. In all the panels (as mentioned in the legend), density contours tracing the densest regions of the thin and thick disc are also shown in red and blue colour, respectively. For the substructures in the halo, linear fits to the distributions are also shown and fit parameters are provided in the legend of the corresponding panels. 
\label{fig:alpha_Fe_disc_halo_components}}
\end{figure*}

\section{Chemical evolution of the thin disc, thick disc, and halo's stellar populations}\label{sec:AlphaFe_DiscHalo}

To find clues about the origins and evolutions of the Galactic disc's and halo’s populations, we search for the differences in their abundance ratios [X/Fe]. In this work, we restrict our analysis to [$\alpha$/Fe]. A run for [$\alpha$/Fe] as a function of [Fe/H] for all the selected samples of the thin disc, thick disc and components of the Galactic halo is shown in Figure \ref{fig:alpha_Fe_disc_halo_components}. For ease in identifying the regions where most of the stars are located, we have shown the number density distributions on a logarithmic scale with the `cubehelix' colour schemes (as shown in the colour bar). The number density distribution is shown for all populations except for the Thamnos and Sequoia due to their relatively smaller samples.
Also, in all the panels (as mentioned in the legends), density contours tracing the densest regions of the thin and thick disc are also shown in red and blue colour, respectively. Additionally, for the substructures in the halo, linear fits to the distributions are also shown, and fit parameters are provided in the legends of the corresponding panels.

As shown in the top panels of Figure \ref{fig:alpha_Fe_disc_halo_components}, the thin and thick disc populations follow the expected evolution of [$\alpha$/Fe] as seen in previous studies like \cite{ReddyLambert2006MNRAS.367.1329R} and references therein. The thin disc stars have the highest abundance ratio [$\alpha$/Fe] in the metal-poor region and follow a decreasing trend in [$\alpha$/Fe] with an increase in [Fe/H]. In the sub-solar metallicity region, however, the distribution flattens. At solar metallicity (i.e., [Fe/H] = 0), the mean [$\alpha$/Fe] in the thin disc is approximately the same as in the Sun, i.e. [$\alpha$/Fe] $\approx$ 0. The metal-poor thin disc stars also have the largest scatter in [$\alpha$/Fe]. 
The thick disc population, on the other hand, has a very distinct [$\alpha$/Fe] distribution compared to the thin disc. The metal-poor thick disc stars (with [Fe/H] $\lessapprox$ $-$1.0) have a mean [$\alpha$/Fe] of about 0.26 dex. For example, the mean [$\alpha$/Fe] for the thick disc stars for the [Fe/H] bin $-$3.0 $\pm$ 0.2, $-$2.5 $\pm$ 0.2, $-$2.0 $\pm$ 0.2, $-$1.5 $\pm$ 0.2, and $-$1.0 $\pm$ 0.2 is 0.33, 0.27, 0.26, 0.26, and 0.25, respectively, with corresponding population SD of 0.24, 0.17, 0.16, 0.15 and 0.11. 
[$\alpha$/Fe] for the thick disc at its peak [Fe/H] of about $-$0.5 is 0.22 dex. In the metal-rich region (with [Fe/H] $>$ $-$0.5), the thick disc follows a decreasing trend in [$\alpha$/Fe] with an increase in [Fe/H] and approach a near thin disc [$\alpha$/Fe] value at super-solar metallicity. The figure also suggests a decrease in the total spread in [$\alpha$/Fe] with an increase in [Fe/H] for the thick disc. At solar metallicity, the mean [$\alpha$/Fe] for the thick disc is slightly higher than the thin disc by about 0.1 dex.

As can be seen from the corresponding panels in Figure \ref{fig:alpha_Fe_disc_halo_components}, the evolution [$\alpha$/Fe] as a function of [Fe/H] for the HD population is similar to the thick disc. However, at any given metallicity, the distribution of the HD is shifted down with respect to the thick disc's distribution (see the position of blue contours) by about 0.1 dex along the [$\alpha$/Fe] axis.

The evolution of Splash is strikingly similar to the thick disc's evolution although only a small fraction of Splash's stars have metallicity higher than $-$0.5 dex. In the metal-poor region, both populations have a flatter distribution of [$\alpha$/Fe] and a linear fit to Splash's data suggests a $\rm \frac{[\alpha/Fe]}{[Fe/H]}$ gradient of $-$0.03 which also passes through the center of thick disc's density contour at its peak metallicity of $-$0.5 dex.

Similar to Splash, the evolution of the Thamnos is also similar to the thick disc, although Thamnos lacks metal-rich stars with [Fe/H] $>$ $-$1.0 dex. At its peak metallicity of $-$1.36 dex, the mean [$\alpha$/Fe] for the Thamnos population is 0.25 dex, which is approximately the same as the mean [$\alpha$/Fe] for the thick disc at this metallicity. Additionally, as shown in Figure \ref{fig:alpha_Fe_disc_halo_components}, a linear fit to the Thamnos data (with gradient $\rm \frac{[\alpha/Fe]}{[Fe/H]}$ of $-$0.014) passes through the center of the thick disc's density contour at its peak metallicity of $-$0.5 dex.

The GE/S, along with the Sequoia and H99, have a distinct distribution of [$\alpha$/Fe] versus [Fe/H] compared to the Galactic disc and other halo components (Figure \ref{fig:alpha_Fe_disc_halo_components}). Unlike the thick disc, HD, Splash and Thamnos, the GE/S, Sequoia and H99 populations follow a decreasing trend in [$\alpha$/Fe] abundance in the metal-poor region also. Linear fits to [$\alpha$/Fe] versus [Fe/H] distributions of these three populations resulted in nearly the same $\frac{[\alpha/Fe]}{[Fe/H]}$ gradient of $-$0.044 which is slightly steeper compared to the thick disc, HD, Splash and Thamnos. This indicates that the GE/S, Sequoia and H99 populations have experienced different chemical evolution than the thick disc, Splash, HD, and Thamnos.

\section{Discussion and concluding remarks}\label{Sec:Conclusion}

When combined, stars' kinematics, age and chemical compositions provide a powerful tool to understand the formation and evolution of a galaxy.
In this study, we use the properties of stars belonging to various kinematically distinct populations of the Galactic disc and halo to find clues about their formation and evolution histories and further to get an insight into the formation and evolution of our home galaxy, Milky Way. For this purpose, we used data from the LAMOST survey, which is one of the largest ground-based spectroscopic surveys in combination with the astrometric and photometric data from the {\it Gaia} survey. Apart from the large sample size, the LAMOST survey also provides good coverage of metal-poor stars with metallicities as low as $-$3.5 dex.
As one of the goals of this work is to understand the formation and evolution of the Galactic halo, which consists of relatively metal-poor stars, the LAMOST survey provides an excellent opportunity to study the Galactic halo.
From the cross-match of the LAMOST and {\it Gaia} catalogues, we kinematically selected samples of the Galactic thin disc, thick disc and halo stars. 

The thin and thick discs (two primary components of the Galactic disc) have moderately symmetric metallicity functions with peak metallicities of $-$0.20 and $-$0.45, respectively (Figure \ref{fig:FeHAgeHist_DiscHalo_allSubStru}). However, their age functions are strikingly different with the thin disc consisting of a relatively younger population (with a peak age of about 5.5 Gyr) while the thick disc primarily consists of an older population (with a peak age of 12.5 Gyr). The age distribution of the thin disc also suggests a smaller secondary peak at an older age of about 12.5 Gyr, which coincides with the peak of the thick disc and also the approximate timeline of the merger GE/S with the Milky Way's progenitors, indicating a possible link between the three. Kinematically and spatially, the thin disc stars are on nearly circular orbits and are localized near the Galactic plane compared to the thick disc stars, the majority of which are on slightly elliptical orbits and located relatively away from the Galactic plane. These results are well in agreement with previous studies like \cite{ReddyLambert2006MNRAS.367.1329R} and references therein.

The CMD for the selected sample of halo stars, as shown in Figure \ref{fig:CMD_Halo}, readily suggest the presence of two colour sequences in the MS as well as the RGB. As discussed in Section \ref{sec:HaloandDiscStarsSelection}, the difference in the metallicities of the halo's two major components (the GE/S and the Splash, also known as the blue and red sequence, respectively) is the reason behind this bi-model CMD for halo. Based on the distribution in the orbital energy versus the angular momentum plane, we selected samples of kinematically distinct stellar populations from the halo. Previous studies like \cite{GallartBernardBrook2019NatAs...3..932G} segregated these two colour sequences based on the distribution in the CMD. However, as shown in Figure \ref{fig:CMD_HaloSubStruct} and discussed in Section \ref{sec:CMD_HaloSubStr}, there is a significant overlap between CMDs of the GE/S and Splash. The CMDs for the other small substructures also overlap with the CMDs for these two major substructures of the halo. These overlaps among the CMDs and possible errors in the estimates of stars' magnitude and colour make it difficult to accurately segregate the halo's substructures purely based on the distribution in the CMD and highlight the importance of the kinematic segregation of these substructures.

Out of our kinematically selected stellar populations/substructures in the galactic halo, the Splash is the largest, followed by the GE/S. The Splash and GE/S have median metallicities of $-$0.75 and $-$1.31, respectively, and are primarily responsible for the appearance of two colour sequences in the halo's CMD (Figure \ref{fig:CMD_HaloSubStruct}). The HD population is the most metal-rich with a mean metallicity of $-$0.43 dex with a population SD of 0.30 dex and consists of both young and old stars. Most of the HD's stars have hotter kinematics than the typical Galactic disc stars and follow slightly elliptical orbits with median orbital eccentricity of 0.35 with small dispersion given by MAD of 0.09 and are also restricted to $|Z|_{\rm max} \lesssim$ 5 kpc (Figure \ref{fig:EccentricityAndZmax_Halo_allSubStru} and Table \ref{table:FeHAgeEccZmax_1}).

All the selected halo sub-structures, except the HD population, are relatively old, with some of their member stars as old as the Universe itself (i.e. about 14 Gyr old). The Splash has a peak age of 13.03 Gyr and is about 1.5 Gyr older than the GE/S, whose peak and median ages are 11.66 and 11.26 Gyr, respectively. Also, most of the Splash and GE/S stars are older than about 8 Gyr. These age estimates are in agreement with many previous studies like \cite{SchusterMoreno2012A&A...538A..21S}, \cite{HawkinsJofre2014MNRAS.445.2575H}, \cite{GeBi2016ApJ...833..161G}, however, they are in contrast to a recent study by \cite{GallartBernardBrook2019NatAs...3..932G} which suggests that both GE/S and Splash (i.e., blue and red colour sequence, respectively) have the same peak age of 13.4 Gyr. For the Splash, our age estimate is in agreement with \cite{GallartBernardBrook2019NatAs...3..932G}'s age estimate. However, for the GE/S, our age estimate is about 2 Gyr lower than the age estimate provided by \cite{GallartBernardBrook2019NatAs...3..932G}. The disagreement between our age estimates (which are in agreement with many previous studies) and \cite{GallartBernardBrook2019NatAs...3..932G} appears to be related to the segregation of these two halo's populations in \cite{GallartBernardBrook2019NatAs...3..932G}. As discussed above and in Section \ref{sec:CMD_HaloSubStr}, the Galactic halo is home to many substructures, of which the Splash and GE/S are the two major ones. The selection of Splash and GE/S samples purely based on the distribution in the CMD is expected to result in large contamination due to overlap in metallicity and the presence of other small substructures in the halo. Our selection of halo's sub-structures is based on stars' kinematics and is independent of their metallicity and distribution in the CMD.

The difference in the peak ages of the Splash and GE/S can be explained based on the environment in which these two halo populations may have formed. The parabolic orbits (with peak orbital eccentricity of about unity) and large $|Z|_{\rm max}$ of GE/S stars hint toward their accreted origin. The GE/S is believed to be a remnant of accreted dwarf galaxy Gaia-Enceladus/Sausage, while the Splash is believed to consist of {\it in situ} formed stars in the Milky Way \citep{HelmiBabusiaux2018Natur.563...85H,GallartBernardBrook2019NatAs...3..932G,KoppelmanHelmi2019A&A...631L...9K}. Based on simulations, studies also suggest an initial mass ratio of about 4:1 between progenitors of the Milky Way and the GE/S. The higher mass galaxies are understood to form stars at a more accelerated pace compared to the low-mass Galaxies \citep[e.g.,][]{KauffmannHeckman2003MNRAS.341...54K,JuneauStephanie2005ApJ...619L.135J,GuglielmoPoggianti2015MNRAS.450.2749G}. Chemically, a higher mass galaxy is also understood to evolve faster compared to a low-mass galaxy \citep{ErbShapley2006ApJ...644..813E}. Splash's higher metallicity but lower peak age compared to the GE/S population supports the idea that the GE/S originated from a less massive progenitor compared to the progenitor of the Milky Way. A sudden drop in age distribution at about 8 Gyr for the GE/S (Figure \ref{fig:FeHAgeHist_DiscHalo_allSubStru}) suggest that the star formation in the GE/S progenitor ceased sometime more than 8 Gyr ago. This may be either due to GE/S progenitor running out of gas to form more new stars or its gas getting stripped off due to interaction with the Milky Way. The age distribution of the Splash population is a lot different than the GE/S. Since its peak age at about 13 Gyr ago, there has been a continuous exponential decay in star count (Figure \ref{fig:FeHAgeHist_DiscHalo_allSubStru}), which can result from Milky Way's gas settling in the disc possibly triggered by the merger of the GE/S progenitor with the Milky Way's progenitor. Also, about 10\% of the Splash's stars are younger than 8 Gyr. \cite{GallartBernardBrook2019NatAs...3..932G}'s sample also confirms the presence of young stars in halo's components \cite[see the left-hand panel of Figure 2 of][]{GallartBernardBrook2019NatAs...3..932G}. The existence of these young stars in the Splash can be explained by the dynamical migration of stars from the Galactic disc and bulge to higher energy orbitals.

The age distribution of the HD sample roughly matches the GE/S's age distribution. Although not very clear due to the small sample size of the HD population, a sudden drop in the HD's age distribution can be seen at about 9 Gyr, which roughly coincides with the sudden drop in the age distribution for the GE/S population. This similarity in the age distribution (especially the drop in the star count around 8 Gyr) along with the disc-like spatial and kinematic features of the HD sample but with relatively energetic orbits than the disc stars suggests that the encounter between the GE/S and the Milky Way may have helped these stars gain higher energy orbits.

Thamnos and Sequoia are the two most metal-poor substructures among our selected substructures in the halo and have peak metallicities of $-$1.36 and $-$1.56, respectively. The Thamnos and Sequoia are believed to be remnants of two small dwarf galaxies which merged with the progenitor of the Milky Way \citep{KoppelmanHelmi2019A&A...631L...9K, MyeongVasiliev2019MNRAS.488.1235M, LimbergRossi2021ApJ...907...10L}. These two populations are also highly retrograde with mean rotational velocities with respect to the LSR of about $-$360 and $-$530 km s$^{-1}$, respectively. Both of these populations have stars with circular to highly elliptical orbits with median eccentricities of 0.49 and 0.51, respectively, with corresponding MAD values of 0.10 and 0.14. Their age distributions are similar to the GE/S's, with very few or no stars younger than 8 Gyr. The higher peak metallicity and older age of the Thamnos compared to the Sequoia is possible if the progenitor of the Thamnos was slightly more massive than the progenitor of the Sequoia, resulting in faster star formation and chemical evolution in the Thamnos compared to the Sequoia.

In conclusion, the lower metallicity and lower peak star formation age of halo's three accreted components (the GE/S, Thamnos and Sequoia) compared to the pro-grade, {\it in situ} components like Splash and HD, favours the $\Lambda$CMD model of galaxy formation in which smaller dwarf galaxies merge with the large complex galaxies like the Milky Way.

\subsection{Insights into the formation of different Galactic components based on [$\alpha$/Fe] -- [Fe/H] distributions}\label{sec:Chap10_XFe_DiscHaloSubStr_discussion}

The amount of $\alpha$-process elements relative to iron in the thin disc, which is also the youngest among the Galactic disc and halo components, follows a decreasing trend with an increase in iron (Figure \ref{fig:alpha_Fe_disc_halo_components}). The exploding high-mass stars as Type II supernovae (SNII) are primary contributors of the $\alpha$-process elements. On the other hand, the exploding low-mass stars as Type Ia supernovae (SNIa) contribute most of the iron peak elements. Hence, the decreasing trend in [$\alpha$/Fe] as a function of an increase in [Fe/H] suggests an increase in the rate of SNIa compared to SNII in the Galactic thin disc. The initial higher [$\alpha$/Fe] in metal-poor thin-disc stars is understood to be due to a higher rate of SNII at earlier times in the inner region of the Galaxy \citep[e.g.][and references therein]{EdvardssonAndersen1993A&A...275..101E,ReddyTomkin2003MNRAS.340..304R}, which resulted in enrichment of ISM with $\alpha$-elements before the star formation in the thin disc peaked about 5 Gyr ago (Figure \ref{fig:FeHAgeHist_DiscHalo_allSubStru}). The presence of a large scatter in [$\alpha$/Fe] abundance for the metal-poor thin-disc stars indicates a poorly mixed gas at the time of formation of these metal-poor stars. Some of these stars, however, may be younger and have originated from the Galactic bulge, which is known to have stars with a wide range in metallicity and $\alpha$-elements abundances. Further studies of other chemical species covering light elements and neutron-capture elements are needed to find the contribution of poorly mixed gas and the Galactic bulge towards the large scatter in [$\alpha$/Fe] abundances in metal-poor thin-disc stars.

Unlike the thin disc, the thick disc has a more complex distribution of [$\alpha$/Fe] abundances (Figure \ref{fig:alpha_Fe_disc_halo_components}).
In the [Fe/H] range from about $-$2.5 to $-$0.5 dex, the number of stars with larger [$\alpha$/Fe] at a given [Fe/H] increases with an increase in [Fe/H], suggesting the dominating contribution of $\alpha$-process elements from the SNII. Also, in this metallicity range, the total spread in [$\alpha$/Fe] at a given [Fe/H] is also increasing with an increase in [Fe/H], suggesting the $\alpha$-elements contribution from SNII was not uniform throughout the Galaxy, and the gas was poorly mixed during the formation of the thick disc. Additionally, the presence of stars with lower [$\alpha$/Fe] further confirms the simultaneous contribution of iron from SNIa in some cases. For [Fe/H] $>$ $-$0.5 dex, the decreasing trend in [$\alpha$/Fe] with respect to an increase in [Fe/H] gives rise to the well-known knee-like distribution for the thick disc and suggests the dominating contribution of iron from SNIa at the later stage of thick disc formation.
These trends are in agreement with previous studies, for example, \cite{EdvardssonAndersen1993A&A...275..101E,ReddyLambert2006MNRAS.367.1329R}. The thin disc is understood to have formed during the merger of a gas-rich system with the progenitor of the Milky Way about 10 Gyr ago \citep[e.g., see][and references therein]{HayesMajewski2018ApJ...852...49H,HaywoodDiMatteo2018ApJ...863..113H,HelmiBabusiaux2018Natur.563...85H,GallartBernardBrook2019NatAs...3..932G,KoppelmanHelmi2019A&A...631L...9K,BelokurovSanders2020MNRAS.494.3880B}. The retrograde component HD and Splash, having [$\alpha$/Fe] abundances similar to the thick disc (except lack of $\alpha$-poor stars), are also believed to have originated during this or a similar merger event \citep{BelokurovSanders2020MNRAS.494.3880B}. The GE/S, the largest accreted component in the Galactic halo, is known to be the remnant of the last major merger in the Milky Way's history and has a progenitor mass of about one-fourth of the mass of Milky Way's progenitor. As shown in Figure \ref{fig:alpha_Fe_disc_halo_components}, the decreasing [$\alpha$/Fe] as a function of [Fe/H] for the GE/S suggests that it GE/S has undergone a distinct chemical evolution than the Milky Way's thick disc and other {\it in situ} components like the HD and Splash. The [$\alpha$/Fe] abundance distribution for the Sequoia and H99 are also similar to the GE/S, suggesting the progenitors of both have undergone similar chemical histories. On the other hand, [$\alpha$/Fe] abundance trend of the Thamnos (which is also believed to be a remnant of a dwarf galaxy) is similar to the Milky Way's {\it in situ} components like HD and Splash (and also metal-poor end of thick disc) and suggests that not all the dwarf galaxies which merged with the Milky Way's progenitors had chemical evolution distinct than the initial chemical evolution of Milky Way progenitor.

To get a more accurate picture of the formation and evolution histories of the Galactic disc and halo and the Galaxy itself, further study of the full spectrum of chemical elements (covering various enrichment channels like $\alpha$-process and neutron-capture process) of these various stellar populations in the Galactic disc and halo is needed. However, the LAMOST's low resolution is not adequate to resolve the small differences present in the compositions of these stellar populations. Hence, further studies based on data from high-resolution spectroscopy surveys should help in finding clues about the formation and evolution histories of the Galaxy and its various stellar populations.

\section*{Acknowledgments} \label{sec:acknowledgments}
\vspace{-1em}
The author is indebted to Prof. Bacham Eswar Reddy (Indian Institute of Astrophysics, Bangalore 560034, India) and Prof. Emeritus David L. Lambert (W.J. McDonald Observatory and Department of Astronomy, The University of Texas at Austin, Austin, TX 78712, USA) for thoughtful discussions, suggestions and comments, which have helped in this study and also resulted in the overall improvement of this manuscript. The author also thanks the anonymous reviewer for valuable suggestions which have helped in the improvement of this manuscript.

This work has made use of data from the European Space Agency (ESA) mission {\it Gaia} (\url{https://www.cosmos.esa.int/gaia}), processed by the {\it Gaia} Data Processing and Analysis Consortium (DPAC, \url{https://www.cosmos.esa.int/web/gaia/dpac/consortium}). Funding for the DPAC has been provided by national institutions, in particular, the institutions participating in the {\it Gaia} Multilateral Agreement.

This work has also made use of data from the LAMOST survey. The data for the LAMOST survey is acquired through the Guoshoujing Telescope. Guoshoujing Telescope (the Large Sky Area Multi-Object Fiber Spectroscopic Telescope; LAMOST) is a National Major Scientific Project built by the Chinese Academy of Sciences. Funding for the project has been provided by the National Development and Reform Commission. LAMOST is operated and managed by the National Astronomical Observatories, Chinese Academy of Sciences.

The initial calculations for this work are carried out using the Delphinus and Fornax servers at the Indian Institute of Astrophysics.

This research has also made use of NASA's Astrophysics Data System.




\bibliographystyle{aasjournal}
\bibliography{ref}

\begin{thebibliography}{}
\expandafter\ifx\csname natexlab\endcsname\relax\def\natexlab#1{#1}\fi
\providecommand{\url}[1]{\href{#1}{#1}}

\bibitem[{{Belokurov} {et~al.}(2020){Belokurov}, {Sanders}, {Fattahi}, {Smith}, {Deason}, {Evans}, \& {Grand}}]{BelokurovSanders2020MNRAS.494.3880B}
{Belokurov}, V., {Sanders}, J.~L., {Fattahi}, A., {et~al.} 2020, \mnras, 494, 3880

\bibitem[{{Bensby} {et~al.}(2003){Bensby}, {Feltzing}, \& {Lundstr{\"o}m}}]{BensbyFeltzing2003A&A...410..527B}
{Bensby}, T., {Feltzing}, S., \& {Lundstr{\"o}m}, I. 2003, \aap, 410, 527

\bibitem[{{Bovy}(2015)}]{JoBovy_galpy2015ApJS..216...29B}
{Bovy}, J. 2015, \apjs, 216, 29

\bibitem[{{Brook} {et~al.}(2020){Brook}, {Kawata}, {Gibson}, {Gallart}, \& {Vicente}}]{BrookKawata2020MNRAS.495.2645B}
{Brook}, C.~B., {Kawata}, D., {Gibson}, B.~K., {Gallart}, C., \& {Vicente}, A. 2020, \mnras, 495, 2645

\bibitem[{{Buder} {et~al.}(2018){Buder}, {Asplund}, {Duong}, {Kos}, {Lind}, {Ness}, {Sharma}, {Bland-Hawthorn}, {Casey}, {de Silva}, {D'Orazi}, {Freeman}, {Lewis}, {Lin}, {Martell}, {Schlesinger}, {Simpson}, {Zucker}, {Zwitter}, {Amarsi}, {Anguiano}, {Carollo}, {Casagrande}, {{\v{C}}otar}, {Cottrell}, {da Costa}, {Gao}, {Hayden}, {Horner}, {Ireland}, {Kafle}, {Munari}, {Nataf}, {Nordlander}, {Stello}, {Ting}, {Traven}, {Watson}, {Wittenmyer}, {Wyse}, {Yong}, {Zinn}, {{\v{Z}}erjal}, \& {Galah Collaboration}}]{BuderAsplund2018MNRAS.478.4513B}
{Buder}, S., {Asplund}, M., {Duong}, L., {et~al.} 2018, \mnras, 478, 4513

\bibitem[{{Buder} {et~al.}(2021){Buder}, {Sharma}, {Kos}, {Amarsi}, {Nordlander}, {Lind}, {Martell}, {Asplund}, {Bland-Hawthorn}, {Casey}, {de Silva}, {D'Orazi}, {Freeman}, {Hayden}, {Lewis}, {Lin}, {Schlesinger}, {Simpson}, {Stello}, {Zucker}, {Zwitter}, {Beeson}, {Buck}, {Casagrande}, {Clark}, {{\v{C}}otar}, {da Costa}, {de Grijs}, {Feuillet}, {Horner}, {Kafle}, {Khanna}, {Kobayashi}, {Liu}, {Montet}, {Nandakumar}, {Nataf}, {Ness}, {Spina}, {Tepper-Garc{\'\i}a}, {Ting}, {Traven}, {Vogrin{\v{c}}i{\v{c}}}, {Wittenmyer}, {Wyse}, {{\v{Z}}erjal}, \& {GALAH Collaboration}}]{BuderSharma2021MNRAS.506..150B}
{Buder}, S., {Sharma}, S., {Kos}, J., {et~al.} 2021, \mnras, 506, 150

\bibitem[{{Buder} {et~al.}(2022){Buder}, {Lind}, {Ness}, {Feuillet}, {Horta}, {Monty}, {Buck}, {Nordlander}, {Bland-Hawthorn}, {Casey}, {de Silva}, {D'Orazi}, {Freeman}, {Hayden}, {Kos}, {Martell}, {Lewis}, {Lin}, {Schlesinger}, {Sharma}, {Simpson}, {Stello}, {Zucker}, {Zwitter}, {Ciuc{\u{a}}}, {Horner}, {Kobayashi}, {Ting}, {Wyse}, \& {Wyse}}]{BuderLind2022MNRAS.510.2407B}
{Buder}, S., {Lind}, K., {Ness}, M.~K., {et~al.} 2022, \mnras, 510, 2407

\bibitem[{{Carollo} {et~al.}(2007){Carollo}, {Beers}, {Lee}, {Chiba}, {Norris}, {Wilhelm}, {Sivarani}, {Marsteller}, {Munn}, {Bailer-Jones}, {Fiorentin}, \& {York}}]{CarolloBeers2007Natur.450.1020C}
{Carollo}, D., {Beers}, T.~C., {Lee}, Y.~S., {et~al.} 2007, \nat, 450, 1020

\bibitem[{{Danielski} {et~al.}(2018){Danielski}, {Babusiaux}, {Ruiz-Dern}, {Sartoretti}, \& {Arenou}}]{DanielskiBabusiaux2018A&A...614A..19D}
{Danielski}, C., {Babusiaux}, C., {Ruiz-Dern}, L., {Sartoretti}, P., \& {Arenou}, F. 2018, \aap, 614, A19

\bibitem[{{Davis} {et~al.}(1985){Davis}, {Efstathiou}, {Frenk}, \& {White}}]{DavisEfstathiou1985ApJ...292..371D}
{Davis}, M., {Efstathiou}, G., {Frenk}, C.~S., \& {White}, S.~D.~M. 1985, \apj, 292, 371

\bibitem[{{De Silva} {et~al.}(2015){De Silva}, {Freeman}, {Bland-Hawthorn}, {Martell}, {de Boer}, {Asplund}, {Keller}, {Sharma}, {Zucker}, {Zwitter}, {Anguiano}, {Bacigalupo}, {Bayliss}, {Beavis}, {Bergemann}, {Campbell}, {Cannon}, {Carollo}, {Casagrande}, {Casey}, {Da Costa}, {D'Orazi}, {Dotter}, {Duong}, {Heger}, {Ireland}, {Kafle}, {Kos}, {Lattanzio}, {Lewis}, {Lin}, {Lind}, {Munari}, {Nataf}, {O'Toole}, {Parker}, {Reid}, {Schlesinger}, {Sheinis}, {Simpson}, {Stello}, {Ting}, {Traven}, {Watson}, {Wittenmyer}, {Yong}, \& {{\v{Z}}erjal}}]{DeSilvaFreeman2015MNRAS.449.2604D}
{De Silva}, G.~M., {Freeman}, K.~C., {Bland-Hawthorn}, J., {et~al.} 2015, \mnras, 449, 2604

\bibitem[{{Edvardsson} {et~al.}(1993){Edvardsson}, {Andersen}, {Gustafsson}, {Lambert}, {Nissen}, \& {Tomkin}}]{EdvardssonAndersen1993A&A...275..101E}
{Edvardsson}, B., {Andersen}, J., {Gustafsson}, B., {et~al.} 1993, \aap, 500, 391

\bibitem[{{Eggen} {et~al.}(1962){Eggen}, {Lynden-Bell}, \& {Sandage}}]{EggenLyndenSandage1962ApJ...136..748E}
{Eggen}, O.~J., {Lynden-Bell}, D., \& {Sandage}, A.~R. 1962, \apj, 136, 748

\bibitem[{{Erb} {et~al.}(2006){Erb}, {Shapley}, {Pettini}, {Steidel}, {Reddy}, \& {Adelberger}}]{ErbShapley2006ApJ...644..813E}
{Erb}, D.~K., {Shapley}, A.~E., {Pettini}, M., {et~al.} 2006, \apj, 644, 813

\bibitem[{{Evans} {et~al.}(2018){Evans}, {Riello}, {De Angeli}, {Carrasco}, {Montegriffo}, {Fabricius}, {Jordi}, {Palaversa}, {Diener}, {Busso}, {Cacciari}, {van Leeuwen}, {Burgess}, {Davidson}, {Harrison}, {Hodgkin}, {Pancino}, {Richards}, {Altavilla}, {Balaguer-N{\'u}{\~n}ez}, {Barstow}, {Bellazzini}, {Brown}, {Castellani}, {Cocozza}, {De Luise}, {Delgado}, {Ducourant}, {Galleti}, {Gilmore}, {Giuffrida}, {Holl}, {Kewley}, {Koposov}, {Marinoni}, {Marrese}, {Osborne}, {Piersimoni}, {Portell}, {Pulone}, {Ragaini}, {Sanna}, {Terrett}, {Walton}, {Wevers}, \& {Wyrzykowski}}]{EvansRiello2018A&A...616A...4E}
{Evans}, D.~W., {Riello}, M., {De Angeli}, F., {et~al.} 2018, \aap, 616, A4

\bibitem[{{Feuillet} {et~al.}(2020){Feuillet}, {Feltzing}, {Sahlholdt}, \& {Casagrande}}]{FeuilletFeltzing2020MNRAS.497..109F}
{Feuillet}, D.~K., {Feltzing}, S., {Sahlholdt}, C.~L., \& {Casagrande}, L. 2020, \mnras, 497, 109

\bibitem[{{Fulbright}(2000)}]{Fulbright2000AJ....120.1841F}
{Fulbright}, J.~P. 2000, \aj, 120, 1841

\bibitem[{{Gaia Collaboration} {et~al.}(2018{\natexlab{a}}){Gaia Collaboration}, {Babusiaux}, {van Leeuwen}, {Barstow}, {Jordi}, {Vallenari}, {Bossini}, {Bressan}, {Cantat-Gaudin}, {van Leeuwen}, {Brown}, {Prusti}, {de Bruijne}, {Bailer-Jones}, {Biermann}, {Evans}, {Eyer}, {Jansen}, {Klioner}, {Lammers}, {Lindegren}, {Luri}, {Mignard}, {Panem}, {Pourbaix}, {Randich}, {Sartoretti}, {Siddiqui}, {Soubiran}, {Walton}, {Arenou}, {Bastian}, {Cropper}, {Drimmel}, {Katz}, {Lattanzi}, {Bakker}, {Cacciari}, {Casta{\~n}eda}, {Chaoul}, {Cheek}, {De Angeli}, {Fabricius}, {Guerra}, {Holl}, {Masana}, {Messineo}, {Mowlavi}, {Nienartowicz}, {Panuzzo}, {Portell}, {Riello}, {Seabroke}, {Tanga}, {Th{\'e}venin}, {Gracia-Abril}, {Comoretto}, {Garcia-Reinaldos}, {Teyssier}, {Altmann}, {Andrae}, {Audard}, {Bellas-Velidis}, {Benson}, {Berthier}, {Blomme}, {Burgess}, {Busso}, {Carry}, {Cellino}, {Clementini}, {Clotet}, {Creevey}, {Davidson}, {De Ridder}, {Delchambre}, {Dell'Oro}, {Ducourant}, {Fern{\'a}ndez-Hern{\'a}ndez},
  {Fouesneau}, {Fr{\'e}mat}, {Galluccio}, {Garc{\'\i}a-Torres}, {Gonz{\'a}lez-N{\'u}{\~n}ez}, {Gonz{\'a}lez-Vidal}, {Gosset}, {Guy}, {Halbwachs}, {Hambly}, {Harrison}, {Hern{\'a}ndez}, {Hestroffer}, {Hodgkin}, {Hutton}, {Jasniewicz}, {Jean-Antoine-Piccolo}, {Jordan}, {Korn}, {Krone-Martins}, {Lanzafame}, {Lebzelter}, {L{\"o}ffler}, {Manteiga}, {Marrese}, {Mart{\'\i}n-Fleitas}, {Moitinho}, {Mora}, {Muinonen}, {Osinde}, {Pancino}, {Pauwels}, {Petit}, {Recio-Blanco}, {Richards}, {Rimoldini}, {Robin}, {Sarro}, {Siopis}, {Smith}, {Sozzetti}, {S{\"u}veges}, {Torra}, {van Reeven}, {Abbas}, {Abreu Aramburu}, {Accart}, {Aerts}, {Altavilla}, {{\'A}lvarez}, {Alvarez}, {Alves}, {Anderson}, {Andrei}, {Anglada Varela}, {Antiche}, {Antoja}, {Arcay}, {Astraatmadja}, {Bach}, {Baker}, {Balaguer-N{\'u}{\~n}ez}, {Balm}, {Barache}, {Barata}, {Barbato}, {Barblan}, {Barklem}, {Barrado}, {Barros}, {Bartholom{\'e} Mu{\~n}oz}, {Bassilana}, {Becciani}, {Bellazzini}, {Berihuete}, {Bertone}, {Bianchi}, {Bienaym{\'e}}, {Blanco-Cuaresma},
  {Boch}, {Boeche}, {Bombrun}, {Borrachero}, {Bouquillon}, {Bourda}, {Bragaglia}, {Bramante}, {Breddels}, {Brouillet}, {Br{\"u}semeister}, {Brugaletta}, {Bucciarelli}, {Burlacu}, {Busonero}, {Butkevich}, {Buzzi}, {Caffau}, {Cancelliere}, {Cannizzaro}, {Carballo}, {Carlucci}, {Carrasco}, {Casamiquela}, {Castellani}, {Castro-Ginard}, {Charlot}, {Chemin}, {Chiavassa}, {Cocozza}, {Costigan}, {Cowell}, {Crifo}, {Crosta}, {Crowley}, {Cuypers}, {Dafonte}, {Damerdji}, {Dapergolas}, {David}, {David}, {de Laverny}, {De Luise}, {De March}, {de Martino}, {de Souza}, {de Torres}, {Debosscher}, {del Pozo}, {Delbo}, {Delgado}, {Delgado}, {Diakite}, {Diener}, {Distefano}, {Dolding}, {Drazinos}, {Dur{\'a}n}, {Edvardsson}, {Enke}, {Eriksson}, {Esquej}, {Eynard Bontemps}, {Fabre}, {Fabrizio}, {Faigler}, {Falc{\~a}o}, {Farr{\`a}s Casas}, {Federici}, {Fedorets}, {Fernique}, {Figueras}, {Filippi}, {Findeisen}, {Fonti}, {Fraile}, {Fraser}, {Fr{\'e}zouls}, {Gai}, {Galleti}, {Garabato}, {Garc{\'\i}a-Sedano}, {Garofalo}, {Garralda},
  {Gavel}, {Gavras}, {Gerssen}, {Geyer}, {Giacobbe}, {Gilmore}, {Girona}, {Giuffrida}, {Glass}, {Gomes}, {Granvik}, {Gueguen}, {Guerrier}, {Guiraud}, {Guti{\'e}}, {Haigron}, {Hatzidimitriou}, {Hauser}, {Haywood}, {Heiter}, {Helmi}, {Heu}, {Hilger}, {Hobbs}, {Hofmann}, {Holland}, {Huckle}, {Hypki}, {Icardi}, {Jan{\ss}en}, {Jevardat de Fombelle}, {Jonker}, {Juh{\'a}sz}, {Julbe}, {Karampelas}, {Kewley}, {Klar}, {Kochoska}, {Kohley}, {Kolenberg}, {Kontizas}, {Kontizas}, {Koposov}, {Kordopatis}, {Kostrzewa-Rutkowska}, {Koubsky}, {Lambert}, {Lanza}, {Lasne}, {Lavigne}, {Le Fustec}, {Le Poncin-Lafitte}, {Lebreton}, {Leccia}, {Leclerc}, {Lecoeur-Taibi}, {Lenhardt}, {Leroux}, {Liao}, {Licata}, {Lindstr{\o}m}, {Lister}, {Livanou}, {Lobel}, {L{\'o}pez}, {Managau}, {Mann}, {Mantelet}, {Marchal}, {Marchant}, {Marconi}, {Marinoni}, {Marschalk{\'o}}, {Marshall}, {Martino}, {Marton}, {Mary}, {Massari}, {Matijevi{\v{c}}}, {Mazeh}, {McMillan}, {Messina}, {Michalik}, {Millar}, {Molina}, {Molinaro}, {Moln{\'a}r}, {Montegriffo},
  {Mor}, {Morbidelli}, {Morel}, {Morris}, {Mulone}, {Muraveva}, {Musella}, {Nelemans}, {Nicastro}, {Noval}, {O'Mullane}, {Ord{\'e}novic}, {Ord{\'o}{\~n}ez-Blanco}, {Osborne}, {Pagani}, {Pagano}, {Pailler}, {Palacin}, {Palaversa}, {Panahi}, {Pawlak}, {Piersimoni}, {Pineau}, {Plachy}, {Plum}, {Poggio}, {Poujoulet}, {Pr{\v{s}}a}, {Pulone}, {Racero}, {Ragaini}, {Rambaux}, {Ramos-Lerate}, {Regibo}, {Reyl{\'e}}, {Riclet}, {Ripepi}, {Riva}, {Rivard}, {Rixon}, {Roegiers}, {Roelens}, {Romero-G{\'o}mez}, {Rowell}, {Royer}, {Ruiz-Dern}, {Sadowski}, {Sagrist{\`a} Sell{\'e}s}, {Sahlmann}, {Salgado}, {Salguero}, {Sanna}, {Santana-Ros}, {Sarasso}, {Savietto}, {Schultheis}, {Sciacca}, {Segol}, {Segovia}, {S{\'e}gransan}, {Shih}, {Siltala}, {Silva}, {Smart}, {Smith}, {Solano}, {Solitro}, {Sordo}, {Soria Nieto}, {Souchay}, {Spagna}, {Spoto}, {Stampa}, {Steele}, {Steidelm{\"u}ller}, {Stephenson}, {Stoev}, {Suess}, {Surdej}, {Szabados}, {Szegedi-Elek}, {Tapiador}, {Taris}, {Tauran}, {Taylor}, {Teixeira}, {Terrett}, {Teyssand
  ier}, {Thuillot}, {Titarenko}, {Torra Clotet}, {Turon}, {Ulla}, {Utrilla}, {Uzzi}, {Vaillant}, {Valentini}, {Valette}, {van Elteren}, {Van Hemelryck}, {Vaschetto}, {Vecchiato}, {Veljanoski}, {Viala}, {Vicente}, {Vogt}, {von Essen}, {Voss}, {Votruba}, {Voutsinas}, {Walmsley}, {Weiler}, {Wertz}, {Wevers}, {Wyrzykowski}, {Yoldas}, {{\v{Z}}erjal}, {Ziaeepour}, {Zorec}, {Zschocke}, {Zucker}, {Zurbach}, \& {Zwitter}}]{GaiaCollaboration_DR2_ObsaHRD2018A&A...616A..10G}
{Gaia Collaboration}, {Babusiaux}, C., {van Leeuwen}, F., {et~al.} 2018{\natexlab{a}}, \aap, 616, A10

\bibitem[{{Gaia Collaboration} {et~al.}(2018{\natexlab{b}}){Gaia Collaboration}, {Brown}, {Vallenari}, {Prusti}, {de Bruijne}, {Babusiaux}, {Bailer-Jones}, {Biermann}, {Evans}, {Eyer}, {Jansen}, {Jordi}, {Klioner}, {Lammers}, {Lindegren}, {Luri}, {Mignard}, {Panem}, {Pourbaix}, {Randich}, {Sartoretti}, {Siddiqui}, {Soubiran}, {van Leeuwen}, {Walton}, {Arenou}, {Bastian}, {Cropper}, {Drimmel}, {Katz}, {Lattanzi}, {Bakker}, {Cacciari}, {Casta{\~n}eda}, {Chaoul}, {Cheek}, {De Angeli}, {Fabricius}, {Guerra}, {Holl}, {Masana}, {Messineo}, {Mowlavi}, {Nienartowicz}, {Panuzzo}, {Portell}, {Riello}, {Seabroke}, {Tanga}, {Th{\'e}venin}, {Gracia-Abril}, {Comoretto}, {Garcia-Reinaldos}, {Teyssier}, {Altmann}, {Andrae}, {Audard}, {Bellas-Velidis}, {Benson}, {Berthier}, {Blomme}, {Burgess}, {Busso}, {Carry}, {Cellino}, {Clementini}, {Clotet}, {Creevey}, {Davidson}, {De Ridder}, {Delchambre}, {Dell'Oro}, {Ducourant}, {Fern{\'a}ndez-Hern{\'a}ndez}, {Fouesneau}, {Fr{\'e}mat}, {Galluccio}, {Garc{\'\i}a-Torres},
  {Gonz{\'a}lez-N{\'u}{\~n}ez}, {Gonz{\'a}lez-Vidal}, {Gosset}, {Guy}, {Halbwachs}, {Hambly}, {Harrison}, {Hern{\'a}ndez}, {Hestroffer}, {Hodgkin}, {Hutton}, {Jasniewicz}, {Jean-Antoine-Piccolo}, {Jordan}, {Korn}, {Krone-Martins}, {Lanzafame}, {Lebzelter}, {L{\"o}ffler}, {Manteiga}, {Marrese}, {Mart{\'\i}n-Fleitas}, {Moitinho}, {Mora}, {Muinonen}, {Osinde}, {Pancino}, {Pauwels}, {Petit}, {Recio-Blanco}, {Richards}, {Rimoldini}, {Robin}, {Sarro}, {Siopis}, {Smith}, {Sozzetti}, {S{\"u}veges}, {Torra}, {van Reeven}, {Abbas}, {Abreu Aramburu}, {Accart}, {Aerts}, {Altavilla}, {{\'A}lvarez}, {Alvarez}, {Alves}, {Anderson}, {Andrei}, {Anglada Varela}, {Antiche}, {Antoja}, {Arcay}, {Astraatmadja}, {Bach}, {Baker}, {Balaguer-N{\'u}{\~n}ez}, {Balm}, {Barache}, {Barata}, {Barbato}, {Barblan}, {Barklem}, {Barrado}, {Barros}, {Barstow}, {Bartholom{\'e} Mu{\~n}oz}, {Bassilana}, {Becciani}, {Bellazzini}, {Berihuete}, {Bertone}, {Bianchi}, {Bienaym{\'e}}, {Blanco-Cuaresma}, {Boch}, {Boeche}, {Bombrun}, {Borrachero},
  {Bossini}, {Bouquillon}, {Bourda}, {Bragaglia}, {Bramante}, {Breddels}, {Bressan}, {Brouillet}, {Br{\"u}semeister}, {Brugaletta}, {Bucciarelli}, {Burlacu}, {Busonero}, {Butkevich}, {Buzzi}, {Caffau}, {Cancelliere}, {Cannizzaro}, {Cantat-Gaudin}, {Carballo}, {Carlucci}, {Carrasco}, {Casamiquela}, {Castellani}, {Castro-Ginard}, {Charlot}, {Chemin}, {Chiavassa}, {Cocozza}, {Costigan}, {Cowell}, {Crifo}, {Crosta}, {Crowley}, {Cuypers}, {Dafonte}, {Damerdji}, {Dapergolas}, {David}, {David}, {de Laverny}, {De Luise}, {De March}, {de Martino}, {de Souza}, {de Torres}, {Debosscher}, {del Pozo}, {Delbo}, {Delgado}, {Delgado}, {Di Matteo}, {Diakite}, {Diener}, {Distefano}, {Dolding}, {Drazinos}, {Dur{\'a}n}, {Edvardsson}, {Enke}, {Eriksson}, {Esquej}, {Eynard Bontemps}, {Fabre}, {Fabrizio}, {Faigler}, {Falc{\~a}o}, {Farr{\`a}s Casas}, {Federici}, {Fedorets}, {Fernique}, {Figueras}, {Filippi}, {Findeisen}, {Fonti}, {Fraile}, {Fraser}, {Fr{\'e}zouls}, {Gai}, {Galleti}, {Garabato}, {Garc{\'\i}a-Sedano}, {Garofalo},
  {Garralda}, {Gavel}, {Gavras}, {Gerssen}, {Geyer}, {Giacobbe}, {Gilmore}, {Girona}, {Giuffrida}, {Glass}, {Gomes}, {Granvik}, {Gueguen}, {Guerrier}, {Guiraud}, {Guti{\'e}rrez-S{\'a}nchez}, {Haigron}, {Hatzidimitriou}, {Hauser}, {Haywood}, {Heiter}, {Helmi}, {Heu}, {Hilger}, {Hobbs}, {Hofmann}, {Holland}, {Huckle}, {Hypki}, {Icardi}, {Jan{\ss}en}, {Jevardat de Fombelle}, {Jonker}, {Juh{\'a}sz}, {Julbe}, {Karampelas}, {Kewley}, {Klar}, {Kochoska}, {Kohley}, {Kolenberg}, {Kontizas}, {Kontizas}, {Koposov}, {Kordopatis}, {Kostrzewa-Rutkowska}, {Koubsky}, {Lambert}, {Lanza}, {Lasne}, {Lavigne}, {Le Fustec}, {Le Poncin-Lafitte}, {Lebreton}, {Leccia}, {Leclerc}, {Lecoeur-Taibi}, {Lenhardt}, {Leroux}, {Liao}, {Licata}, {Lindstr{\o}m}, {Lister}, {Livanou}, {Lobel}, {L{\'o}pez}, {Managau}, {Mann}, {Mantelet}, {Marchal}, {Marchant}, {Marconi}, {Marinoni}, {Marschalk{\'o}}, {Marshall}, {Martino}, {Marton}, {Mary}, {Massari}, {Matijevi{\v{c}}}, {Mazeh}, {McMillan}, {Messina}, {Michalik}, {Millar}, {Molina}, {Molinaro},
  {Moln{\'a}r}, {Montegriffo}, {Mor}, {Morbidelli}, {Morel}, {Morris}, {Mulone}, {Muraveva}, {Musella}, {Nelemans}, {Nicastro}, {Noval}, {O'Mullane}, {Ord{\'e}novic}, {Ord{\'o}{\~n}ez-Blanco}, {Osborne}, {Pagani}, {Pagano}, {Pailler}, {Palacin}, {Palaversa}, {Panahi}, {Pawlak}, {Piersimoni}, {Pineau}, {Plachy}, {Plum}, {Poggio}, {Poujoulet}, {Pr{\v{s}}a}, {Pulone}, {Racero}, {Ragaini}, {Rambaux}, {Ramos-Lerate}, {Regibo}, {Reyl{\'e}}, {Riclet}, {Ripepi}, {Riva}, {Rivard}, {Rixon}, {Roegiers}, {Roelens}, {Romero-G{\'o}mez}, {Rowell}, {Royer}, {Ruiz-Dern}, {Sadowski}, {Sagrist{\`a} Sell{\'e}s}, {Sahlmann}, {Salgado}, {Salguero}, {Sanna}, {Santana-Ros}, {Sarasso}, {Savietto}, {Schultheis}, {Sciacca}, {Segol}, {Segovia}, {S{\'e}gransan}, {Shih}, {Siltala}, {Silva}, {Smart}, {Smith}, {Solano}, {Solitro}, {Sordo}, {Soria Nieto}, {Souchay}, {Spagna}, {Spoto}, {Stampa}, {Steele}, {Steidelm{\"u}ller}, {Stephenson}, {Stoev}, {Suess}, {Surdej}, {Szabados}, {Szegedi-Elek}, {Tapiador}, {Taris}, {Tauran}, {Taylor},
  {Teixeira}, {Terrett}, {Teyssandier}, {Thuillot}, {Titarenko}, {Torra Clotet}, {Turon}, {Ulla}, {Utrilla}, {Uzzi}, {Vaillant}, {Valentini}, {Valette}, {van Elteren}, {Van Hemelryck}, {van Leeuwen}, {Vaschetto}, {Vecchiato}, {Veljanoski}, {Viala}, {Vicente}, {Vogt}, {von Essen}, {Voss}, {Votruba}, {Voutsinas}, {Walmsley}, {Weiler}, {Wertz}, {Wevers}, {Wyrzykowski}, {Yoldas}, {{\v{Z}}erjal}, {Ziaeepour}, {Zorec}, {Zschocke}, {Zucker}, {Zurbach}, \& {Zwitter}}]{BrownGaiaDr2Summary2018}
{Gaia Collaboration}, {Brown}, A.~G.~A., {Vallenari}, A., {et~al.} 2018{\natexlab{b}}, \aap, 616, A1

\bibitem[{{Gallart} {et~al.}(2019){Gallart}, {Bernard}, {Brook}, {Ruiz-Lara}, {Cassisi}, {Hill}, \& {Monelli}}]{GallartBernardBrook2019NatAs...3..932G}
{Gallart}, C., {Bernard}, E.~J., {Brook}, C.~B., {et~al.} 2019, Nature Astronomy, 3, 932

\bibitem[{{Ge} {et~al.}(2016){Ge}, {Bi}, {Chen}, {Li}, {Zhao}, {Liu}, {Ferguson}, \& {Wu}}]{GeBi2016ApJ...833..161G}
{Ge}, Z.~S., {Bi}, S.~L., {Chen}, Y.~Q., {et~al.} 2016, \apj, 833, 161

\bibitem[{{Gratton} {et~al.}(2003){Gratton}, {Carretta}, {Desidera}, {Lucatello}, {Mazzei}, \& {Barbieri}}]{GrattonCarretta2003A&A...406..131G}
{Gratton}, R.~G., {Carretta}, E., {Desidera}, S., {et~al.} 2003, \aap, 406, 131

\bibitem[{{Green}(2011)}]{Green2011BASI...39..289G}
{Green}, D.~A. 2011, Bulletin of the Astronomical Society of India, 39, 289

\bibitem[{{Guglielmo} {et~al.}(2015){Guglielmo}, {Poggianti}, {Moretti}, {Fritz}, {Calvi}, {Vulcani}, {Fasano}, \& {Paccagnella}}]{GuglielmoPoggianti2015MNRAS.450.2749G}
{Guglielmo}, V., {Poggianti}, B.~M., {Moretti}, A., {et~al.} 2015, \mnras, 450, 2749

\bibitem[{{Hawkins} {et~al.}(2014){Hawkins}, {Jofr{\'e}}, {Gilmore}, \& {Masseron}}]{HawkinsJofre2014MNRAS.445.2575H}
{Hawkins}, K., {Jofr{\'e}}, P., {Gilmore}, G., \& {Masseron}, T. 2014, \mnras, 445, 2575

\bibitem[{{Hayes} {et~al.}(2018){Hayes}, {Majewski}, {Shetrone}, {Fern{\'a}ndez-Alvar}, {Allende Prieto}, {Schuster}, {Carigi}, {Cunha}, {Smith}, {Sobeck}, {Almeida}, {Beers}, {Carrera}, {Fern{\'a}ndez-Trincado}, {Garc{\'\i}a-Hern{\'a}ndez}, {Geisler}, {Lane}, {Lucatello}, {Matthews}, {Minniti}, {Nitschelm}, {Tang}, {Tissera}, \& {Zamora}}]{HayesMajewski2018ApJ...852...49H}
{Hayes}, C.~R., {Majewski}, S.~R., {Shetrone}, M., {et~al.} 2018, \apj, 852, 49

\bibitem[{{Haywood} {et~al.}(2018){Haywood}, {Di Matteo}, {Lehnert}, {Snaith}, {Khoperskov}, \& {G{\'o}mez}}]{HaywoodDiMatteo2018ApJ...863..113H}
{Haywood}, M., {Di Matteo}, P., {Lehnert}, M.~D., {et~al.} 2018, \apj, 863, 113

\bibitem[{{Helmi}(2008)}]{Helmi2008A&ARv..15..145H}
{Helmi}, A. 2008, \aapr, 15, 145

\bibitem[{{Helmi} {et~al.}(2018){Helmi}, {Babusiaux}, {Koppelman}, {Massari}, {Veljanoski}, \& {Brown}}]{HelmiBabusiaux2018Natur.563...85H}
{Helmi}, A., {Babusiaux}, C., {Koppelman}, H.~H., {et~al.} 2018, \nat, 563, 85

\bibitem[{{Helmi} {et~al.}(1999){Helmi}, {White}, {de Zeeuw}, \& {Zhao}}]{HelmiWhite1999Natur.402...53H}
{Helmi}, A., {White}, S. D.~M., {de Zeeuw}, P.~T., \& {Zhao}, H. 1999, \nat, 402, 53

\bibitem[{{Horta} {et~al.}(2021){Horta}, {Schiavon}, {Mackereth}, {Pfeffer}, {Mason}, {Kisku}, {Fragkoudi}, {Allende Prieto}, {Cunha}, {Hasselquist}, {Holtzman}, {Majewski}, {Nataf}, {O'Connell}, {Schultheis}, \& {Smith}}]{HortaSchiavon2021MNRAS.500.1385H}
{Horta}, D., {Schiavon}, R.~P., {Mackereth}, J.~T., {et~al.} 2021, \mnras, 500, 1385

\bibitem[{{Ishigaki} {et~al.}(2010){Ishigaki}, {Chiba}, \& {Aoki}}]{IshigakiChiba2010PASJ...62..143I}
{Ishigaki}, M., {Chiba}, M., \& {Aoki}, W. 2010, \pasj, 62, 143

\bibitem[{{Ishigaki} {et~al.}(2021){Ishigaki}, {Hartwig}, {Tarumi}, {Leung}, {Tominaga}, {Kobayashi}, {Magg}, {Simionescu}, \& {Nomoto}}]{IshigakiHartwig2021MNRAS.506.5410I}
{Ishigaki}, M.~N., {Hartwig}, T., {Tarumi}, Y., {et~al.} 2021, \mnras, 506, 5410

\bibitem[{{Jonsell} {et~al.}(2005){Jonsell}, {Edvardsson}, {Gustafsson}, {Magain}, {Nissen}, \& {Asplund}}]{JonsellEdvardsson2005A&A...440..321J}
{Jonsell}, K., {Edvardsson}, B., {Gustafsson}, B., {et~al.} 2005, \aap, 440, 321

\bibitem[{{Jordi} {et~al.}(2010){Jordi}, {Gebran}, {Carrasco}, {de Bruijne}, {Voss}, {Fabricius}, {Knude}, {Vallenari}, {Kohley}, \& {Mora}}]{JordiGaiaPhotometry2010}
{Jordi}, C., {Gebran}, M., {Carrasco}, J.~M., {et~al.} 2010, \aap, 523, A48

\bibitem[{{Juneau} {et~al.}(2005){Juneau}, {Glazebrook}, {Crampton}, {McCarthy}, {Savaglio}, {Abraham}, {Carlberg}, {Chen}, {Le Borgne}, {Marzke}, {Roth}, {J{\o}rgensen}, {Hook}, \& {Murowinski}}]{JuneauStephanie2005ApJ...619L.135J}
{Juneau}, S., {Glazebrook}, K., {Crampton}, D., {et~al.} 2005, \apjl, 619, L135

\bibitem[{{Kauffmann} {et~al.}(2003){Kauffmann}, {Heckman}, {White}, {Charlot}, {Tremonti}, {Peng}, {Seibert}, {Brinkmann}, {Nichol}, {SubbaRao}, \& {York}}]{KauffmannHeckman2003MNRAS.341...54K}
{Kauffmann}, G., {Heckman}, T.~M., {White}, S. D.~M., {et~al.} 2003, \mnras, 341, 54

\bibitem[{{Kim} {et~al.}(2002){Kim}, {Demarque}, {Yi}, \& {Alexand er}}]{KimDemarque2002Y2Isochrones}
{Kim}, Y.-C., {Demarque}, P., {Yi}, S.~K., \& {Alexand er}, D.~R. 2002, \apjs, 143, 499

\bibitem[{{Klement}(2010)}]{Klement2010A&ARv..18..567K}
{Klement}, R.~J. 2010, \aapr, 18, 567

\bibitem[{{Koppelman} {et~al.}(2019){Koppelman}, {Helmi}, {Massari}, {Price-Whelan}, \& {Starkenburg}}]{KoppelmanHelmi2019A&A...631L...9K}
{Koppelman}, H.~H., {Helmi}, A., {Massari}, D., {Price-Whelan}, A.~M., \& {Starkenburg}, T.~K. 2019, \aap, 631, L9

\bibitem[{{Limberg} {et~al.}(2021){Limberg}, {Rossi}, {Beers}, {Perottoni}, {P{\'e}rez-Villegas}, {Santucci}, {Abuchaim}, {Placco}, {Lee}, {Christlieb}, {Norris}, {Bessell}, {Ryan}, {Wilhelm}, {Rhee}, \& {Frebel}}]{LimbergRossi2021ApJ...907...10L}
{Limberg}, G., {Rossi}, S., {Beers}, T.~C., {et~al.} 2021, \apj, 907, 10

\bibitem[{{Lindegren} {et~al.}(2018){Lindegren}, {Hern{\'a}ndez}, {Bombrun}, {Klioner}, {Bastian}, {Ramos-Lerate}, {de Torres}, {Steidelm{\"u}ller}, {Stephenson}, {Hobbs}, {Lammers}, {Biermann}, {Geyer}, {Hilger}, {Michalik}, {Stampa}, {McMillan}, {Casta{\~n}eda}, {Clotet}, {Comoretto}, {Davidson}, {Fabricius}, {Gracia}, {Hambly}, {Hutton}, {Mora}, {Portell}, {van Leeuwen}, {Abbas}, {Abreu}, {Altmann}, {Andrei}, {Anglada}, {Balaguer-N{\'u}{\~n}ez}, {Barache}, {Becciani}, {Bertone}, {Bianchi}, {Bouquillon}, {Bourda}, {Br{\"u}semeister}, {Bucciarelli}, {Busonero}, {Buzzi}, {Cancelliere}, {Carlucci}, {Charlot}, {Cheek}, {Crosta}, {Crowley}, {de Bruijne}, {de Felice}, {Drimmel}, {Esquej}, {Fienga}, {Fraile}, {Gai}, {Garralda}, {Gonz{\'a}lez-Vidal}, {Guerra}, {Hauser}, {Hofmann}, {Holl}, {Jordan}, {Lattanzi}, {Lenhardt}, {Liao}, {Licata}, {Lister}, {L{\"o}ffler}, {Marchant}, {Martin-Fleitas}, {Messineo}, {Mignard}, {Morbidelli}, {Poggio}, {Riva}, {Rowell}, {Salguero}, {Sarasso}, {Sciacca}, {Siddiqui}, {Smart},
  {Spagna}, {Steele}, {Taris}, {Torra}, {van Elteren}, {van Reeven}, \& {Vecchiato}}]{LindegrenHernandez2018A&A...616A...2L}
{Lindegren}, L., {Hern{\'a}ndez}, J., {Bombrun}, A., {et~al.} 2018, \aap, 616, A2

\bibitem[{{Mackereth} \& {Bovy}(2018)}]{MackerethBovy2018PASP..130k4501M}
{Mackereth}, J.~T., \& {Bovy}, J. 2018, \pasp, 130, 114501

\bibitem[{{Majewski} {et~al.}(2017){Majewski}, {Schiavon}, {Frinchaboy}, {Allende Prieto}, {Barkhouser}, {Bizyaev}, {Blank}, {Brunner}, {Burton}, {Carrera}, {Chojnowski}, {Cunha}, {Epstein}, {Fitzgerald}, {Garc{\'\i}a P{\'e}rez}, {Hearty}, {Henderson}, {Holtzman}, {Johnson}, {Lam}, {Lawler}, {Maseman}, {M{\'e}sz{\'a}ros}, {Nelson}, {Nguyen}, {Nidever}, {Pinsonneault}, {Shetrone}, {Smee}, {Smith}, {Stolberg}, {Skrutskie}, {Walker}, {Wilson}, {Zasowski}, {Anders}, {Basu}, {Beland}, {Blanton}, {Bovy}, {Brownstein}, {Carlberg}, {Chaplin}, {Chiappini}, {Eisenstein}, {Elsworth}, {Feuillet}, {Fleming}, {Galbraith-Frew}, {Garc{\'\i}a}, {Garc{\'\i}a-Hern{\'a}ndez}, {Gillespie}, {Girardi}, {Gunn}, {Hasselquist}, {Hayden}, {Hekker}, {Ivans}, {Kinemuchi}, {Klaene}, {Mahadevan}, {Mathur}, {Mosser}, {Muna}, {Munn}, {Nichol}, {O'Connell}, {Parejko}, {Robin}, {Rocha-Pinto}, {Schultheis}, {Serenelli}, {Shane}, {Silva Aguirre}, {Sobeck}, {Thompson}, {Troup}, {Weinberg}, \& {Zamora}}]{MajewskiSchiavon2017AJ....154...94M}
{Majewski}, S.~R., {Schiavon}, R.~P., {Frinchaboy}, P.~M., {et~al.} 2017, \aj, 154, 94

\bibitem[{{McMillan}(2017)}]{McMillan2017MNRAS.465...76M}
{McMillan}, P.~J. 2017, \mnras, 465, 76

\bibitem[{{Myeong} {et~al.}(2019){Myeong}, {Vasiliev}, {Iorio}, {Evans}, \& {Belokurov}}]{MyeongVasiliev2019MNRAS.488.1235M}
{Myeong}, G.~C., {Vasiliev}, E., {Iorio}, G., {Evans}, N.~W., \& {Belokurov}, V. 2019, \mnras, 488, 1235

\bibitem[{{Nissen} \& {Schuster}(1997)}]{NissenSchuster1997A&A...326..751N}
{Nissen}, P.~E., \& {Schuster}, W.~J. 1997, \aap, 326, 751

\bibitem[{{Nissen} \& {Schuster}(2010)}]{NissenSchuster2010A&A...511L..10N}
---. 2010, \aap, 511, L10

\bibitem[{{Nissen} \& {Schuster}(2011)}]{NissenSchuster2011A&A...530A..15N}
---. 2011, \aap, 530, A15

\bibitem[{{Ojha}(2001)}]{Ojha2001MNRAS.322..426O}
{Ojha}, D.~K. 2001, \mnras, 322, 426

\bibitem[{{Perryman} {et~al.}(1997){Perryman}, {Lindegren}, {Kovalevsky}, {Hoeg}, {Bastian}, {Bernacca}, {Cr{\'e}z{\'e}}, {Donati}, {Grenon}, {Grewing}, {van Leeuwen}, {van der Marel}, {Mignard}, {Murray}, {Le Poole}, {Schrijver}, {Turon}, {Arenou}, {Froeschl{\'e}}, \& {Petersen}}]{PerrymanLindegren1997A&A...323L..49P}
{Perryman}, M.~A.~C., {Lindegren}, L., {Kovalevsky}, J., {et~al.} 1997, \aap, 323, L49

\bibitem[{{Press} \& {Schechter}(1974)}]{PressSchechter1974ApJ...187..425P}
{Press}, W.~H., \& {Schechter}, P. 1974, \apj, 187, 425

\bibitem[{{Queiroz} {et~al.}(2020){Queiroz}, {Anders}, {Chiappini}, {Khalatyan}, {Santiago}, {Steinmetz}, {Valentini}, {Miglio}, {Bossini}, {Barbuy}, {Minchev}, {Minniti}, {Garc{\'\i}a Hern{\'a}ndez}, {Schultheis}, {Beaton}, {Beers}, {Bizyaev}, {Brownstein}, {Cunha}, {Fern{\'a}ndez-Trincado}, {Frinchaboy}, {Lane}, {Majewski}, {Nataf}, {Nitschelm}, {Pan}, {Roman-Lopes}, {Sobeck}, {Stringfellow}, \& {Zamora}}]{QueirozAnders2020A&A...638A..76Q}
{Queiroz}, A.~B.~A., {Anders}, F., {Chiappini}, C., {et~al.} 2020, \aap, 638, A76

\bibitem[{{Ram{\'\i}rez} {et~al.}(2014){Ram{\'\i}rez}, {Mel{\'e}ndez}, {Bean}, {Asplund}, {Bedell}, {Monroe}, {Casagrande}, {Schirbel}, {Dreizler}, {Teske}, {Tucci Maia}, {Alves-Brito}, \& {Baumann}}]{RamirezMelendezBean2014A&A...572A..48R}
{Ram{\'\i}rez}, I., {Mel{\'e}ndez}, J., {Bean}, J., {et~al.} 2014, \aap, 572, A48

\bibitem[{{Reddy} {et~al.}(2006){Reddy}, {Lambert}, \& {Allende Prieto}}]{ReddyLambert2006MNRAS.367.1329R}
{Reddy}, B.~E., {Lambert}, D.~L., \& {Allende Prieto}, C. 2006, \mnras, 367, 1329

\bibitem[{{Reddy} {et~al.}(2003){Reddy}, {Tomkin}, {Lambert}, \& {Allende Prieto}}]{ReddyTomkin2003MNRAS.340..304R}
{Reddy}, B.~E., {Tomkin}, J., {Lambert}, D.~L., \& {Allende Prieto}, C. 2003, \mnras, 340, 304

\bibitem[{{Robin} {et~al.}(1996){Robin}, {Haywood}, {Creze}, {Ojha}, \& {Bienayme}}]{RobinHaywoodCreze1996A&A...305..125R}
{Robin}, A.~C., {Haywood}, M., {Creze}, M., {Ojha}, D.~K., \& {Bienayme}, O. 1996, \aap, 305, 125

\bibitem[{{Salaris} {et~al.}(1993){Salaris}, {Chieffi}, \& {Straniero}}]{Salaris1993ApJ...414..580S}
{Salaris}, M., {Chieffi}, A., \& {Straniero}, O. 1993, \apj, 414, 580

\bibitem[{{Schuster} {et~al.}(2012){Schuster}, {Moreno}, {Nissen}, \& {Pichardo}}]{SchusterMoreno2012A&A...538A..21S}
{Schuster}, W.~J., {Moreno}, E., {Nissen}, P.~E., \& {Pichardo}, B. 2012, \aap, 538, A21

\bibitem[{{Searle} \& {Zinn}(1978)}]{SearleZinn1978ApJ...225..357S}
{Searle}, L., \& {Zinn}, R. 1978, \apj, 225, 357

\bibitem[{{Stephens} \& {Boesgaard}(2002)}]{StephensBoesgaard2002AJ....123.1647S}
{Stephens}, A., \& {Boesgaard}, A.~M. 2002, \aj, 123, 1647

\bibitem[{{Xiang} {et~al.}(2019){Xiang}, {Ting}, {Rix}, {Sandford}, {Buder}, {Lind}, {Liu}, {Shi}, \& {Zhang}}]{XiangTingRix2019ApJS..245...34X}
{Xiang}, M., {Ting}, Y.-S., {Rix}, H.-W., {et~al.} 2019, \apjs, 245, 34

\bibitem[{{Yi} {et~al.}(2001){Yi}, {Demarque}, {Kim}, {Lee}, {Ree}, {Lejeune}, \& {Barnes}}]{YiDemarque2001Y2Isochrones}
{Yi}, S., {Demarque}, P., {Kim}, Y.-C., {et~al.} 2001, \apjs, 136, 417

\bibitem[{{Zhao} {et~al.}(2012){Zhao}, {Zhao}, {Chu}, {Jing}, \& {Deng}}]{ZhaoZhao2012RAA....12..723Z}
{Zhao}, G., {Zhao}, Y.-H., {Chu}, Y.-Q., {Jing}, Y.-P., \& {Deng}, L.-C. 2012, Research in Astronomy and Astrophysics, 12, 723

\end{thebibliography}



\end{document}